\documentclass[prb,twocolumn,amsmath,amssymb,floatfix]{revtex4}
\usepackage{graphicx}
\usepackage{bm}
\usepackage{amsmath,amssymb}
\usepackage{dcolumn}
\usepackage{bm}
\usepackage{epsfig}
\usepackage{longtable}
\usepackage{subfigure}
\usepackage{multirow}
\usepackage{xcolor}

\newcommand{\bfr}{ {\bf r}} 
\newcommand{\bfrp}{ {\bf r'}} 
\newcommand{\dv}{ {\Delta \hat{v}}} 

\newcommand{\GDV}{\textsc{GAUSSIAN}}

\begin{document}

\title{ Renormalized Second-order Perturbation Theory for the Electron Correlation Energy: Concept, Implementation, and Benchmarks}
\author{Xinguo Ren,$^1$ Patrick Rinke,$^1$ Gustavo E. Scuseria,$^2$ Matthias Scheffler$^1$}
\affiliation{$^1$Fritz-Haber-Institut der Max-Planck-Gesellschaft,
Faradayweg 4-6, 14195, Berlin, Germany \\
$^2$Department of Chemistry and Department of Physics $\&$ Astronomy, Rice University, Houston, Texas 77005, USA
}

\begin{abstract}
We present a renormalized second-order perturbation theory (rPT2), based on
a Kohn-Sham (KS) reference state, for the electron correlation energy that includes 
the random-phase approximation (RPA), second-order screened exchange (SOSEX),
and renormalized single excitations (rSE). These three terms all
involve a summation of certain types of diagrams to infinite order, and
can be viewed as ``\textit{renormalization}" of the 2nd-order direct, exchange, and 
single excitation (SE) terms of Rayleigh-Schr\"odinger perturbation
theory based on an KS reference.  In this work we establish the concept of rPT2 and present 
the numerical details of our SOSEX and rSE implementations. 
A preliminary version of rPT2, in which the renormalized SE (rSE) contribution was treated 
approximately, has already been benchmarked 
for molecular atomization energies and chemical reaction barrier heights and
shows a well balanced performance [Paier \textit{et al}, New J. Phys. \textbf{14}, 043002 (2012)]. 
In this work, we present a refined version of rPT2, in which we evaluate the rSE series of diagrams
rigorously. We then extend the benchmark studies 
to non-covalent interactions, including the rare-gas dimers,
and the S22 and S66 test sets. Despite some remaining shortcomings, we conclude that rPT2 gives an 
overall satisfactory performance across different chemical environments, and is a promising step 
towards a generally applicable electronic structure approach. 
\end{abstract}

\maketitle

\section{Introduction}
Density-functional theory (DFT) \cite{Hohenberg/Kohn:1964,Kohn/Sham:1965} has played a significant role in first-principles electronic-structure calculations in physics, chemistry, materials science, and biophysics over the 
past decades.  
DFT offers an in principle exact formalism for computing ground-state energies of electronic systems,
but in practice the exchange-correlation (XC) energy functional has to be approximated. Existing 
approximations to the XC functional can be classified into different rungs according to a hierarchical
scheme known as ``Jacob's ladder".\cite{Perdew/Schmidt:2001}  The random-phase 
approximation (RPA), \cite{Bohm/Pines:1953,Gell-Mann/Brueckner:1957} which in the context of DFT 
\cite{Langreth/Perdew:1977,Gunnarsson/Lundqvist:1976} amounts to treating the exchange energy exactly 
and the 
correlation energy at the level of RPA, is on the fifth and highest rung of this ladder. RPA has received considerable attention (for two recent reviews, see Refs.~\onlinecite{Eshuis/Bates/Furche:2012} and \onlinecite{Ren/etal:2012b}) since its first application to realistic systems.\cite{Furche:2001} This is largely due to the fact that RPA has shown great promise in resolving difficulties encountered by the local-density and generalized gradient approximations (LDA/GGAs) to DFT.  The resolution of the ``CO adsorption puzzle",
\cite{Feibelman:2001,Ren/Rinke/Scheffler:2009,Schimka/etal:2010} the encouraging behavior for
the ``strongly correlated" $f$-electron Ce metal, \cite{Casadei/etal:2012} and the excellent performance of
RPA (and its variants) across a wide range of systems including solids, 
\cite{Harl/Kresse:2008,Harl/Kresse:2009,Schimka/etal:2010}
van der Waals (vdW) bonded molecules, \cite{Janesko/Henderson/Scuseria:2009,Janesko/Henderson/Scuseria:2009b,Toulouse/etal:2009,Ren/etal:2011,Lu/Li/Rocca/Galli:2009} 
and thermochemistry \cite{Eshuis/Furche:2011} are just a few examples.

Quantitatively, however, RPA itself does not always provide the desired accuracy. It was found empirically that
 the common practice of evaluating both the exact-exchange and the RPA correlation energy in a post-processing way using Kohn-Sham (KS) or generalized KS orbitals leads to a systematic underestimation of bond strengths in both molecules and solids.\cite{Furche:2001,Schimka/etal:2010,Paier/etal:2010,Ren/etal:2011} Iterating RPA to self-consistency does not alleviate this problem. \cite{Caruso/etal:2012}
Various attempts have been made in the past to improve the standard RPA scheme,
\cite{Yan/Perdew/Kurth:2000,Toulouse/etal:2009,Janesko/Henderson/Scuseria:2009,Grueneis/etal:2009,Paier/etal:2010,Hesselmann:2011,Hesselmann/Goerling:2011b,Ren/etal:2011,Ruzsinszky/etal:2011,Angyan/etal:2011,Olsen/Thygesen:2012} with varying degrees of success. Here we will focus on two flavors of beyond-RPA schemes that both alleviate the underbinding problem of RPA:  the 
second-order screened exchange (SOSEX) \cite{Freeman:1977,Grueneis/etal:2009,Paier/etal:2010} and 
the single-excitation (SE) correction. \cite{Ren/etal:2011}  SOSEX was originally formulated in the context of coupled cluster theory,\cite{Freeman:1977,Grueneis/etal:2009} and accounts for the antisymmetric nature of the many-electron wave function. Like RPA, it can be interpreted as an infinite summation of a set of topologically similar diagrams. 
\cite{Goldstone:1957,Grueneis/etal:2009,Ren/etal:2012b} Adding SOSEX to RPA makes the theory one-electron ``self-correlation" free. The SE correction, on the other hand, accounts for the fact that the KS orbitals are not optimal 
for a post-processing perturbation treatment at the exact-exchange level.\cite{Ren/etal:2011} 
In analogy to RPA and SOSEX, one can also identify a 
sequence of single excitation diagrams. Summing these
to infinite order yields what we called the \textit{renormalized single-excitation} (rSE) contribution \cite{Ren/etal:2011} to the electron correlation energy.
Combining all three contributions -- RPA, SOSEX, and rSE -- leads to the ``RPA+SOSEX+rSE" 
scheme, or as we shall refer to it in this work: \textit{renormalized 2nd-order perturbation theory}, in short 
rPT2 (note 
that in Ref.~[\onlinecite{Ren/etal:2012b}] we used the acronym r2PT). The name is inspired by second-order Rayleigh-Schr{\"o}dinger perturbation theory (RSPT) that becomes \emph{renormalized} through the infinite summations. This can be compared to the commonly used second-order M{\o}ller-Plesset (MP2)
method, which is a straight (bare) second-order RSPT based on the Hartree-Fock reference.

A preliminary version of rPT2, in which an approximate treatment of rSE was invoked, had been 
benchmarked for atomization energies of molecules and chemical reaction barrier heights 
in Ref.~\onlinecite{Paier/etal:2012}. We found that rPT2 gives the ``most balanced" performance compared
to other RPA-based schemes.
However, this approximate treatment of rSE turns out to be problematic for weak interactions and exhibits an
unphysical behavior in, e.g., the binding energy curve of rare gas dimers. 
In this work, we will show how a rigorous evaluation of rSE can be carried out. From here on, rPT2 will refer to this revised scheme and not the approximate  version presented in Ref.~\onlinecite{Paier/etal:2012}.  We will, in 
particular, examine the performance of rPT2 for weakly bonded molecules, 
including rare-gas dimers, and the widely used S22 and S66 test sets of Hobza and co-authors. 
\cite{Jurecka/etal:2006,Rezac/etal:2008,Rezac/etal:2011} For completeness, we will also revisit
the benchmark sets for the G2 atomization energies of Curtiss \textit{et al.} \cite{Curtiss/etal:1997} 
and the chemical reaction barrier heights of Truhlar and co-authors \cite{Zhao/Nuria/Truhlar:2005,Zhao/Truhlar:2006} 
for which the performance of the preliminary
rPT2 version was first tested in Ref.~\onlinecite{Paier/etal:2012}. In addition to the concept of 
rPT2 and benchmark studies, we will also present a different
way of formulating the SOSEX term, that corresponds to the adiabatic connection formulation of 
SOSEX (AC-SOSEX) by Jansen, Liu, and {\'A}nyg{\'a}n (JLA),\cite{Jansen/etal:2010} and that reflects 
our actual implementation.
Our benchmark studies show that rPT2 represents an overall improvement over RPA, and gives a gratifying 
performance across different electronic and chemical environments. We also identify remaining shortcomings 
that will guide further developments of the theory.

The remainder of the paper is organized as follows:  In Sec.~\ref{sec:theory}, the basic theory and implementation of rPT2 is presented. This is followed by a systematic benchmark test for rPT2 for a range of systems in Sec.~\ref{sec:results}. Conclusions are drawn in Sec.~\ref{sec:conc}. Further details of our implementation and derivations 
will be given in Appendices.

\section{\label{sec:theory}Theory}

In this section the theoretical foundation of rPT2 will be presented. We first recapitulate the basics of the
 RPA+SOSEX method in Sec.~\ref{sec:th_rpa+sosex}, and present the theory in a way that reflects  its 
implementation in the Fritz Haber Institute \emph{ab initio molecular simulations} (FHI-aims) code package.\cite{Blum/etal:2009,Ren/etal:2012} This is followed by the derivation of an algebraic expression for the rSE term -- the third ingredient in rPT2.  A discussion of the underlying physics behind the rPT2 
method is then presented from a diagrammatic point of view in Sec.~\ref{sec:th_rPT2}.

\subsection{\label{sec:th_rpa+sosex}The RPA+SOSEX method}
The RPA method can be formulated in different ways (for a review, see Ref.~\onlinecite{Eshuis/Bates/Furche:2012} and \onlinecite{Ren/etal:2012b}). In the DFT context, RPA can be derived from the adiabatic-connection fluctuation-dissipation (ACFD) theorem, 
\cite{Langreth/Perdew:1977,Gunnarsson/Lundqvist:1976}
 whereby the RPA correlation energy is expressed as
 \begin{align}
      E_\text{c}^\text{RPA}  = - & \frac{1}{2\pi} \int_0^1 d\lambda \int_0^\infty d \omega \iint d\bfr d\bfrp
         v(\bfr,\bfrp)   \times \nonumber \\
        &   \left[\chi_\lambda^\text{RPA}(\bfrp, \bfr, i\omega) - \chi_0(\bfrp, \bfr, i\omega)\right] \, .
  \label{Eq:RPA-ACFD}
 \end{align}
$\chi_0(i\omega)$ is the KS independent-particle density-response function 
 \begin{equation}
      \chi_0(\bfr,\bfrp,i\omega) = \sum_{ia} \left[
             \frac{\psi_i^\ast(\bfr)\psi_a(\bfr)\psi_i(\bfrp)\psi_a(\bfrp)}{\epsilon_i-\epsilon_a-i\omega} + c.c.
          \right]
  \label{Eq:chi_0_realspace}
 \end{equation}
where $\psi_{i,a}(\bfr)$ and $\epsilon_{i,a}$ are the KS single-particle orbitals and orbital energies, 
and $c.c.$ the ``complex conjugate".  Here and in the following we adopt
the following convention: $i,j$ correspond to occupied and $a,b$ to unoccupied (or virtual) spin orbitals, 
whereas $p,q$ apply to general cases.
$\chi_\lambda^\text{RPA}(i\omega)$ in Eq.~(\ref{Eq:RPA-ACFD}) is the RPA response function 
of a fictitious system with a scaled Coulomb interaction $\frac{\lambda}{|\bfr - \bfrp|}$ (with $0 \le \lambda \le 1$), and satisfies the Dyson equation
 \begin{equation}
   \chi_\lambda = \chi_0 + \chi_0 \lambda v \chi_\lambda \, .
   \label{Eq:RPA_Dyson}
 \end{equation}
Representing $\chi_0$ and $v$ in the ``particle-hole basis"
$\{ \psi_i^\ast(\bfr)\psi_a(\bfr), \psi_a^\ast(\bfr)\psi_i(\bfr) \}$, one can obtain the RPA correlation energy 
by solving the following eigenvalue problem
\cite{Furche:2001}
  \begin{equation}
     \left( \begin{array}{rr}  A & B^\ast \\
                              -B & -A^\ast 
            \end{array} 
     \right)
     \left( \begin{array}{c}  X_n \\
                              Y_n
            \end{array} 
     \right)
     = \left( \begin{array}{c} X_n \\ Y_n 
                \end{array} \right) \omega_n  \, ,
    \label{Eq:RPA_eigen}
  \end{equation}
where $A_{ia,jb}=(\epsilon_a-\epsilon_i)\delta_{ij}\delta_{ab}+\langle ib|aj\rangle$, and 
$B_{ia,jb}=\langle ij|ab\rangle$. The two-electron Coulomb integrals are 
 \begin{equation}
   \langle pq|rs \rangle = \iint dx_1 dx_2 
           \frac{\psi_p^\ast(x_1)\psi_r(x_1)\psi_q^\ast(x_2)\psi_s(x_2)}{|\bfr_1-\bfr_2|} \, ,
   \label{Eq:2eri}
 \end{equation}
where $x=(\bfr,\sigma)$ is a combined space-spin variable.
As demonstrated by Furche,\cite{Furche:2008} after solving Eq.~(\ref{Eq:RPA_eigen}), the RPA correlation energy 
can be written as
 \begin{equation}
   E_\text{c}^\text{RPA} = \frac{1}{2}\text{Tr} (\omega -A) = \frac{1}{2} \left[ {\sum_n}^\prime \omega_n - 
               \sum_{ia} A_{ia,ia} \right] \, ,
  \label{Eq:RPA_plasmon}
 \end{equation}
where $\sum_n^\prime$ implies that the summation over $n$ is restricted to positive eigenvalues $\omega_n$. 

\begin{figure*}
\centering
\includegraphics[width=0.6\textwidth,clip]{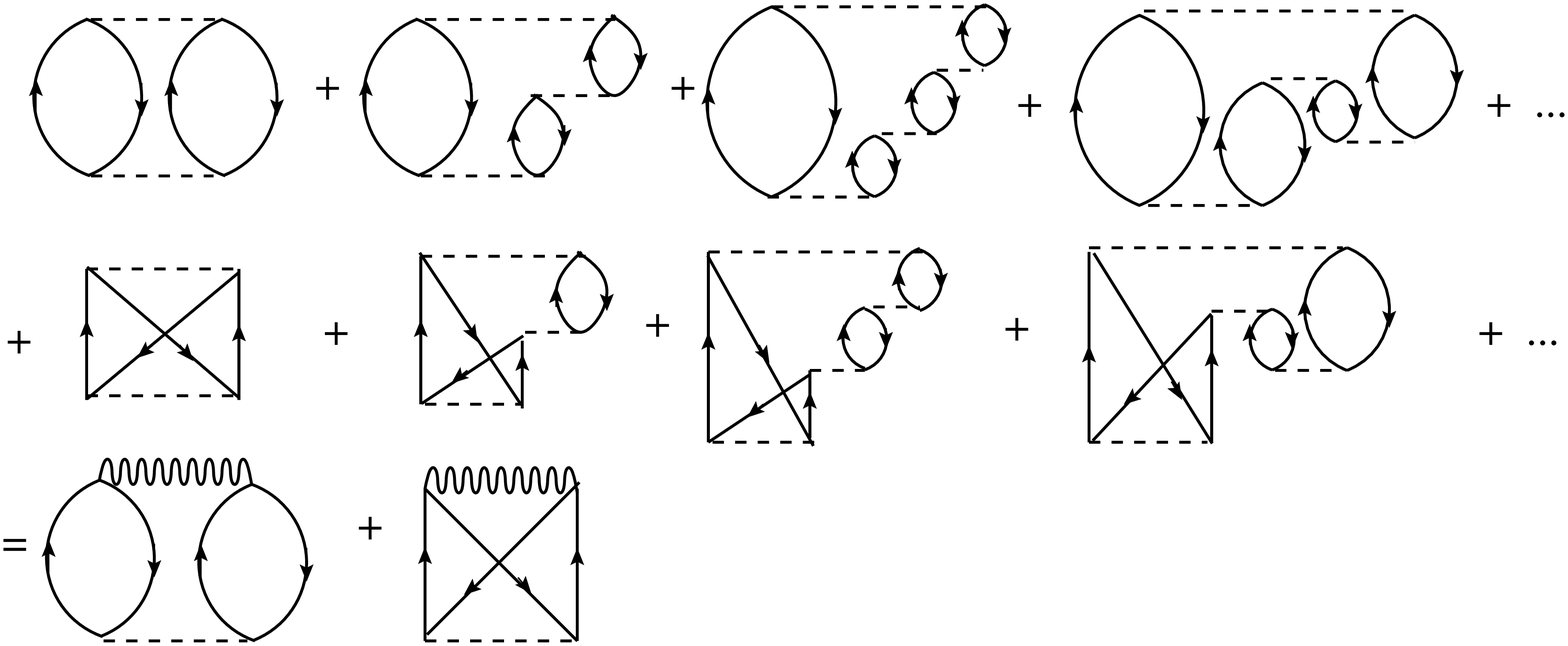}
\caption{Goldstone diagrams for RPA (first row) and SOSEX (second row) contributions. 
         Dashed lines represent bare Coulomb interactions, and full lines correspond to KS electrons (arrow up) 
         and holes (arrow down). Third row: RPA+SOSEX energy in the coupled-cluster
        context. The wiggly together with the arrowed, solid lines represent the direct ring-CCD amplitudes 
        $T_{ia,jb}$ (see Eq.~(\ref{Eq:EcSOSEX_CCD})). The contraction between the direct ring-CCD amplitudes and the
        bare Coulomb interactions (dashed lines) yields the RPA+SOSEX correlation energy. }
\label{Fig:rpa+sosex_diagrams}
\end{figure*}
Scuseria \textit{et al.} demonstrated that an equivalent formulation of the RPA correlation energy of
Eq.~(\ref{Eq:RPA_plasmon}) can be obtained from an approximate coupled-cluster doubles (CCD) theory 
\cite{Scuseria/Henderson/Sorensen:2008} in which only the ``ring diagrams" are kept (see the first row
of Fig.~\ref{Fig:rpa+sosex_diagrams}). In the CCD theory, only double excitation contributions are
included in the ``cluster operator" which generates the interacting many-body ground-state wavefunction
through the exponential ansatz. By contrast, in the more often used CCSD approach, both single and double excitations
are included.  Within the CCD formulation of RPA, the
key quantities are the (\emph{direct}) ring-CCD amplitudes $T_{ia,jb}$, which (in the case of
real canonical spin orbitals) are determined by the following Riccati equation,
 \begin{equation}
   B + AT + TA + TBT = 0\, .
  \label{Eq:Riccati}
 \end{equation}
Due to the quadratic nature of this equation, one should take care to ensure that the physical 
solution is taken.\cite{Henderson/Scuseria:2010} 
The RPA correlation energy in this ring-CCD formulation is then given by
  \begin{equation}
      E_\text{c}^\text{RPA} = \frac{1}{2} \text{Tr} (BT) = \frac{1}{2}\sum_{ij,ab} \langle ij|ab \rangle T_{jb,ia}
     \label{Eq:RPA_CCD}\, .
  \end{equation}
We note that this is often called \emph{direct} RPA in the quantum chemistry literature to emphasize the fact that
higher-order exchange-type contributions are not included. 

Now the RPA+SOSEX correlation energy can be conveniently introduced 
\cite{Freeman:1977,Grueneis/etal:2009,Paier/etal:2010} by antisymmetrizing the Coulomb integral in 
Eq.~(\ref{Eq:RPA_CCD}), \cite{Grueneis/etal:2009}
  \begin{equation}
      E_\text{c}^\text{RPA+SOSEX} = \frac{1}{2} \text{Tr} (\tilde{B}T) = \frac{1}{2}
         \sum_{ij,ab} \langle ij||ab \rangle T_{jb,ia}\, ,
     \label{Eq:EcSOSEX_CCD}
  \end{equation}
where $\tilde{B}_{ia,jb}=\langle ij||ab \rangle = \langle ij|ab\rangle - \langle ij|ba\rangle $.
The SOSEX correction term  itself is
  \begin{equation}
      E_\text{c}^\text{SOSEX} = (\tilde{B}T) = -\frac{1}{2} \sum_{ij,ab} \langle ij|ba \rangle T_{jb,ia}\, .
     \label{Eq:EcSOSEX_alone_CCD}
  \end{equation}
Physically, the SOSEX correction introduces higher-order exchange processes that can also be represented by an infinite summation of  Goldstone diagrams (see the second row of Fig.~\ref{Fig:rpa+sosex_diagrams}).  
This infinite summation is condensed into the ring-CCD amplitudes whose contraction with the bare Coulomb interaction
(after antisymmetrization) yields the RPA+SOSEX correlation energy as illustrated by the third-row diagrams 
in Fig.~\ref{Fig:rpa+sosex_diagrams}.

In a coupled cluster code the SOSEX energy can be readily computed once the direct ring-CCD amplitudes $T_{ia,jb}$ are available. A slightly different variant of SOSEX can be obtained in the ACFD framework, as 
shown by JLA \cite{Jansen/etal:2010}. We will show later, that although not identical, these two SOSEX formulations produce very similar results. Our implementation in the FHI-aims code  \cite{Blum/etal:2009,Ren/etal:2012} follows  the ACFD route. To illustrate our approach let us first present an alternative way to Eq.~(\ref{Eq:RPA-ACFD}) of expressing the RPA correlation energy within ACFD
before we introduce the corresponding SOSEX extension.
Eq.~(\ref{Eq:RPA_Dyson}) yields
 \begin{align}
  \chi_\lambda^\text{RPA}(i\omega) = & \chi_0(i\omega) + \chi_0(i\omega) \lambda v 
                       \chi_\lambda^\text{RPA}(i\omega) \nonumber \\
                     = & \chi_0(i\omega) + \lambda \chi_0(i\omega) v \chi_0(i\omega) +  \nonumber \\
                  &  \lambda^2 \chi_0(i\omega) v \chi_0(i\omega) v \chi_0(i\omega) + \cdots \, .
  \label{Eq:RPA_response}
 \end{align}
The RPA correlation energy in Eq.~(\ref{Eq:RPA-ACFD}) can then be rewritten as
 \begin{widetext}
 \begin{eqnarray}
      E_\text{c}^\text{RPA} & = & -  \frac{1}{2\pi} \int_0^1 d\lambda \int_0^\infty d\omega 
         \text{Tr} \left[ \chi_0(i\omega)v \chi_0(i\omega)\cdot \lambda v
        +  \chi_0(i\omega)v \chi_0(i\omega)\cdot \lambda^2 v\chi_0(i\omega)v + \cdots \right]\,  
          \label{Eq:RPA_geometry_series} \\
   &= &-\frac{1}{2\pi} \int_0^1 d\lambda \int_0^\infty d\omega  \text{Tr} \left[ \chi_0(i\omega)v \chi_0(i\omega)
          W_\lambda(i\omega) \right] \nonumber \\
   & =& -\frac{1}{2\pi} 
          \int_0^\infty d\omega  \text{Tr} \left[ \chi_0(i\omega)v \chi_0(i\omega)
          \bar{W}(i\omega) \right] \, , \label{Eq:EcRPA_new} 
 \end{eqnarray}
 \end{widetext}
where 
\begin{equation}
W_\lambda(i\omega) = \lambda v/(1 - \lambda\chi_0(i\omega)v)\, 
 \label{Eq:W_lambda}
\end{equation}
is the coupling-constant-dependent screened Coulomb interaction and 
\begin{equation}
 \bar{W}(i\omega) = \int_0^1 d\lambda W_\lambda(i\omega)
 \label{Eq:W_bar}
\end{equation}
the coupling-constant-averaged screened Coulomb interaction. In this context we would like to point out that the 
first diagram in the third row 
of  Fig.~\ref{Fig:rpa+sosex_diagrams} can alternatively be interpreted as the pictorial representation of 
equation~(\ref{Eq:EcRPA_new}).  
Now the bubbles correspond to $\chi_0$, dashed lines to the bare Coulomb interaction, and wiggly lines to the
corresponding screened interaction $\bar{W}(i\omega)$. 

Expressing $\chi_0$ again in terms of the ``particle-hole basis" (defined below Eq.~(\ref{Eq:RPA_Dyson})) and using Eq~(\ref{Eq:chi_0_realspace}), Eq.~(\ref{Eq:EcRPA_new}) can be recast into
\begin{widetext}
\begin{align}
E_c^\text{RPA}  = 
     & \frac{1}{2\pi} \int_0^\infty d\omega \sum_{ia,jb} \left[ 
           \frac{\langle aj | ib \rangle \langle ib| \bar{W}(i\omega)|aj\rangle} 
           {(\epsilon_i-\epsilon_a-i\omega)(\epsilon_j-\epsilon_b-i\omega)} + 
           \frac{\langle ab | ij \rangle \langle ij| \bar{W}(i\omega)|ab\rangle} 
           {(\epsilon_i-\epsilon_a-i\omega)(\epsilon_j-\epsilon_b + i\omega)} \right. \nonumber \\ 
     & \hspace{2.4cm}  \left. 
           \frac{\langle ij | ab \rangle \langle ab| \bar{W}(i\omega)|ij\rangle} 
           {(\epsilon_i-\epsilon_a+i\omega)(\epsilon_j-\epsilon_b-i\omega)} + 
           \frac{\langle ib | aj \rangle \langle aj| \bar{W}(i\omega)|ib\rangle} 
           {(\epsilon_i-\epsilon_a+i\omega)(\epsilon_j-\epsilon_b + i\omega)} \right] \, , 
\end{align}
\end{widetext}
where $\langle pq| \bar{W}(i\omega)|rs\rangle$ is defined in analogy to $\langle pq||rs\rangle$  
in Eq.~(\ref{Eq:2eri}),
by replacing the bare Coulomb interaction $v$ by the screened (and frequency-dependent) one, $\bar{W}(i\omega)$.

For real canonical spin orbitals we find $\langle aj|ib\rangle = \langle ib|aj\rangle = \langle ab|ij\rangle = 
\langle ij|ab\rangle$. The same relations 
hold for the screened Coulomb repulsion integrals. The above equation then simplifies to
\begin{align}
  E_c^\text{RPA}  = \frac{1}{2\pi} \int_0^\infty d\omega & \sum_{ia,jb}  \langle ij|ab \rangle 
       \langle ij |\bar{W}(i\omega)| ab\rangle \times \nonumber \\
             &  {\cal F }_{ia}(i\omega) {\cal F }_{jb}(i\omega) \,
 \label{Eq:EcRPA_simpl}
\end{align}
with the factors 
\begin{equation}
 {\cal F }_{ia}(i\omega)=2(\epsilon_i-\epsilon_a)/[(\epsilon_i-\epsilon_a)^2+\omega^2] \, .
 \label{Eq:factor}
\end{equation}

Now, in analogy to the (\emph{direct}) ring-CCD formulation of SOSEX in Eq.~(\ref{Eq:EcSOSEX_alone_CCD}), one can
obtain a corresponding SOSEX term (the so-called ``AC-SOSEX") from Eq.~(\ref{Eq:EcRPA_simpl}), by exchanging the ``$a,b$" indices in 
$\langle ij|ba \rangle$ (with an additional minus sign),
 \begin{align}
   E_c^\text{AC-SOSEX} = -\frac{1}{2\pi} \int_0^\infty d\omega & \sum_{ia,jb} \langle ij|ba \rangle 
       \langle ij |\bar{W}(i\omega)| ab\rangle \times \nonumber \\
        &  {\cal F }_{ia}(i\omega) {\cal F }_{jb}(i\omega) \, .
  \label{Eq:EcSOSEX_simpl}
 \end{align}
Then, using the resolution-of-identity technique,\cite{Dunlap/Connolly/Sabin:1979,Feyereisen/Fitzgerald/Komornicki:1993,Weigend/Haser/Patzelt/Ahlrichs:1998,Ren/etal:2012} Eq.~(\ref{Eq:EcSOSEX_simpl}) can be implemented with relative ease.
The implementation details of Eq.~(\ref{Eq:EcSOSEX_simpl}) in FHI-aims are presented in 
Appendix \ref{sec:appendix}. 

To make closer contact with the expression given in Ref.~\onlinecite{Jansen/etal:2010}, we note that 
Eq.~(\ref{Eq:EcRPA_simpl}) can be further rewritten:
 \begin{equation}
      E_c^\text{AC-SOSEX}  = - \frac{1}{2} \sum_{ia,jb} \langle ij|ba \rangle  \bar{P}_{ia,jb}\, ,
      \label{Eq:EcSOSEX_DM}
 \end{equation}
where
 \begin{equation}
   \bar{P}_{ia,jb} = \frac{1}{\pi}  \int_0^\infty d\omega \langle ij |\bar{W}(i\omega)| ab\rangle {\cal F }_{ia}(i\omega) {\cal F }_{jb}(i\omega)\,  
  \label{Eq:2pDM}
 \end{equation}
is the coupling-strength averaged (two-particle) density matrix.

As shown by JLA\cite{Jansen/etal:2010}, 
Eq.~(\ref{Eq:EcSOSEX_DM}) is usually not identical to the original ring-CCD based SOSEX in 
Eq.~(\ref{Eq:EcSOSEX_alone_CCD}) (except for one- and two-electron cases). However, the difference between them is very small (relative difference in RPA+SOSEX correlation energy less that $0.15\%$), 
as first noted in Ref.~\onlinecite{Angyan/etal:2011} and also confirmed here. 
In table~\ref{Tab:SOSEX} we present the 
RPA and SOSEX correlation energies ($E_\text{c}^\text{RPA}$ and $E_\text{c}^\text{SOSEX}$), as well as the 
RPA and RPA+SOSEX atomization energies for five molecules. The vanishingly small  differences in the RPA energies are due to the different implementations in FHI-aims and the development version of the \GDV\cite{gdv-g1} code (e.g., FHI-aims employs
the RI approximation and treats the Gaussian orbitals numerically). The 
difference in the SOSEX and AC-SOSEX correlation energies reflects the intrinsic differences of the two SOSEX formulations. Nevertheless, the differences are very small and have little practical importance,
in particular for atomization energies.
\begin{widetext}
\begin{table*}
  \caption{\label{Tab:SOSEX} RPA and SOSEX (total) correlation energies (in Hartree), as well as
   RPA and RPA+SOSEX atomization energies (in kcal/mol) 
   for five molecules.  The ``AC-SOSEX" numbers are computed using FHI-aims based on 
   Eq.~(\ref{Eq:EcSOSEX_simpl}), whereas the original ring-CCD based SOSEX numbers are computed
   using a development version of the \GDV\cite{gdv-g1} suite of programs.
    All calculations were done with Gaussian cc-pVQZ basis set and frozen-core ($1s$) approximation.
    The reference orbitals are obtained using the GGA functional constructed by Perdew, Burke, and
    Ernzerhof (PBE).\cite{Perdew/Burke/Ernzerhof:1996} Note that in the upper part of the table 
    only the RPA or (AC-)SOSEX correlation contribution is included, whereas in the lower part 
    the numbers are obtained from the total energy (including also the Hartree-Fock part) differences. }
   \begin{tabular*}{0.95\textwidth}{@{\extracolsep{\fill}}l@{\hspace{5mm}}cccccccc}
  \hline\hline \\[-2ex]
    \multicolumn{8}{c}{Correlation energy (Hartree)} \\[1ex]
   \hline \\[-2ex]
    & \multicolumn{3}{c}{RPA} & & \multicolumn{3}{c}{AC-SOSEX/SOSEX}  \\[0.8ex]
     \cline{2-4}   \cline{5-8} \\[-2ex]
    & FHI-aims & \GDV & difference & & FHI-aims & \GDV & difference \\
    &          &          &            & & (AC-SOSEX) & (SOSEX) & \\[0.8ex]
     \cline{2-4}   \cline{5-8}  \\[-2ex]
  CO   &   -0.593778   &     -0.593786  &    ~0.000008 &  & 0.218954   &    0.217977   &  0.000977 \\[0.3ex]
  N$_2$   &   -0.606368   &     -0.606391  &    ~0.000023 &  & 0.224069   &    0.222955   &  0.001114 \\[0.3ex]
  O$_2$   &   -0.730348   &     -0.730364  &    ~0.000016 &  &0.283384    &    0.281073   &  0.002311 \\[0.3ex]
  CH$_4$  &   -0.381735   &     -0.381730  &    -0.000005 &  & 0.155242   &    0.154933   &  0.000309 \\[0.3ex]
  C$_2$H$_2$ &   -0.539435   &     -0.539439  &    ~0.000006 &  & 0.207348   &    0.206514   &  0.000834 \\[0.5ex]
   \hline \\[-2ex]
    \multicolumn{8}{c}{Atomization energy (kcal/mol)} \\[1ex]
   \hline \\[-2ex]
    & \multicolumn{3}{c}{RPA} & & \multicolumn{3}{c}{RPA+AC-SOSEX/RPA+SOSEX}  \\[0.8ex]
     \cline{2-4}   \cline{5-8} \\[-2ex]
    & FHI-aims & \GDV & difference & & FHI-aims & \GDV & difference \\
    &          &          &            & & (AC-SOSEX) & (SOSEX) & \\[0.8ex]
     \cline{2-4}   \cline{5-8}  \\[-2ex]
  CO   & 239.16  &   239.18   &  -0.02  &  & 246.88  &  246.86   &  ~0.02 \\[0.3ex]
  N$_2$   & 217.58  &   217.59   &  -0.01  &  & 209.24  &  209.10   &  ~0.14  \\[0.3ex]
  O$_2$   & 108.02  &   108.03   &  -0.01  &  & ~98.11  &  ~98.71   & -0.60 \\[0.3ex]
  CH$_4$  & 400.15  &   400.13   &  ~0.02  &  & 415.33  &  415.29   &  ~0.04 \\[0.3ex]
  C$_2$H$_2$ & 373.43  &   373.45   &  -0.02  &  & 391.63  &  391.71   & -0.08 \\[0.5ex]
  \hline\hline
\end{tabular*}
\end{table*}
\end{widetext}

Our benchmark results presented in section~\ref{sec:results} are based on the AC-SOSEX scheme. 
However, since the numerical difference between the two SOSEX flavors are very small, our 
conclusion should also apply to the original ring-CCD based SOSEX.


\subsection{\label{sec:th_rSE}The rSE correction and the semi-canonicalization method}
In Ref. \onlinecite{Ren/etal:2011}, we showed that it is advantageous to complement the RPA correlation energy 
with a 
correction term arising from single excitations. The single excitation correction derives directly from Rayleigh-Schr\"{o}dinger perturbation theory (RSPT) and adopts a simple form in terms of the
single-particle orbitals 
 \begin{equation}
   E^\text{SE}_\text{c} = \sum_{ia}\frac{|\langle \psi_i |\hat{f}| \psi_a \rangle|^2}{\epsilon_i-\epsilon_a}
                        = \sum_{ia} \frac{|f_{ia}|^2}{\epsilon_i-\epsilon_a}\, .
   \label{Eq:SE}
 \end{equation}
Here $|\psi_{i(a)}\rangle$ and $\epsilon_{i(a)}$ refer to occupied (unoccupied) Kohn-Sham (KS) orbitals and
the corresponding orbital energies. $\hat{f}$ is the 
single-particle Hartree-Fock (HF) Hamiltonian, or the so-called Fock operator. We have presented the derivation of Eq.~(\ref{Eq:SE}) already in Ref.~\onlinecite{Ren/etal:2011}, but include it here for completeness in Appendix~\ref{app:SE_deriv}.  Denoting  the single-particle KS Hamiltonian $\hat{h}^0$, we obtain 
$\langle \psi_i |\hat{f}| \psi_a \rangle = \langle \psi_i |\hat{h}^0+\Delta \hat{v}| \psi_a \rangle =
\langle \psi_i |\Delta \hat{v}| \psi_a \rangle $ when $\psi_i$, $\psi_a$ are eigenfunctions of $\hat{h}^0$.
$\dv$ is the difference between the HF exact-exchange potential and the KS exchange-correlation
potential.  A similar SE contribution is encountered in the context of KS density functional perturbation 
theory. \cite{Goerling/Levy:1993,Bartlett:2010,Jiang/Engel:2006} However, we emphasize that here we followed the 
procedure of RSPT  to derive
Eq.~(\ref{Eq:SE}), instead of the ACFD formalism, which requires the electron-density to be fixed along the 
adiabatic-connection path. Whether the two procedures will yield significantly different results is
a subject of further studies.

\begin{figure*}
\centering
\includegraphics[width=0.6\textwidth,clip]{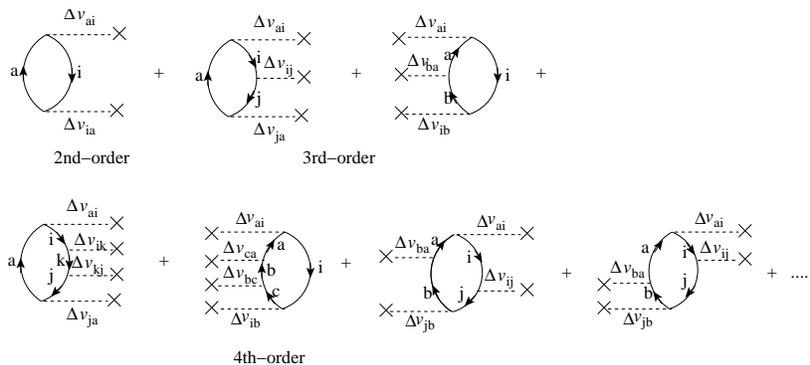}
\caption{Goldstone diagrams for a sequence of correlation-energy terms arising from single excitations. Summing
 these up to infinite order yields the \emph{renormalized} single excitation (rSE) contribution. Here $\Delta v_{pq}=\langle\psi_p |\hat{f}-\hat{h}^0|\psi_q\rangle$, and note $\Delta v_{ia} = f_{ia}$. }
\label{Fig:rSE_diagrams}
\end{figure*}
From the viewpoint of RSPT, Eq.~(\ref{Eq:SE}) represents a second-order correlation energy. As such it suffers from the same divergence problem as 2nd-order M{\o}ller-Plesset perturbation theory for metallic systems when the single-particle KS gap closes. A remedy suggested in 
Ref.~\onlinecite{Ren/etal:2011} was to follow the RPA spirit and to sum a sequence of higher-order SE terms to infinite order. Such higher-order SE terms can also be represented in terms of Goldstone
diagrams, as illustrated in Fig.~\ref{Fig:rSE_diagrams}. We refer to this infinite summation of SE terms as  \emph{renormalized} single excitations (rSE) as alluded to in the introduction.

The influence of the rSE correction was first examined in  Ref.~\onlinecite{Paier/etal:2012}, albeit in an 
approximate way. 
There a so-called ``diagonal" approximation to rSE (denoted here as ``rSE-diag") was used, in which  only  
terms with $``i=j=k=\cdots"$ and $``a=b=c=\cdots"$ were included. The remaining ``off-diagonal" terms were 
omitted. A similar approximation has been used in summing up the Epstein-Nesbet ladder-type diagrams
in Ref.~\onlinecite{Jiang/Engel:2006}.  In this way, the sequence of diagrams falls into a geometrical series. 
Summing them up yields the following simple expression
 \begin{equation}
   E^\text{rSE-diag}_\text{c} = \sum_{ia}\frac{|f_{ia}|^2}
   {\epsilon_i-\epsilon_a + \Delta v_{ii} -  \Delta v_{aa}}\, ,
   \label{Eq:rSE}
 \end{equation}
where $\Delta v_{pq} = \langle \psi_p | \dv | \psi_q \rangle$.
The additional term $\Delta v_{ii} -  \Delta v_{aa}$ that appears
in the denominator is negative definite and removes the divergence problem
even for vanishing KS gaps. The addition of rSE-diag
to RPA and RPA+SOSEX has been benchmarked
for atomization energies and reaction barriers in Ref.~\onlinecite{Paier/etal:2012}.
We found that the renormalization (i.e., going from SE to rSE-diag) has a tendency to slightly reduce atomization energies, but the overall effect is not significant. For chemical reaction barrier heights, on the other hand, the renormalization is crucial for the transition states, that typically have a rather small energy gap.

The diagonal approximation in Eq.~(\ref{Eq:rSE}) is not invariant under unitary transformations in the space of occupied and  unoccupied orbitals. More importantly, however, it  can lead to an unphysical behavior in the potential-energy surface of weakly interacting systems,  as will be shown in Sec.~\ref{sec:rare-gas-dimer}. Recently we discovered that it is straightforward to include  the ``off-diagonal" elements as well, and to treat the rSE term rigorously. 
In Appendix~\ref{sec:app/rSE} we 
illustrate in detail how the infinite summation of the diagrams
depicted in Fig.~\ref{Fig:rSE_diagrams} can be carried out. Here we only present the key steps that 
lead to the final expression, and that are needed in practical calculations. 

First, the occupied and unoccupied blocks of the Fock matrix (evaluated with KS orbitals) need to be constructed
  \begin{align}
    f_{ij} & = \langle \psi_i|\hat{f}|\psi_j \rangle = \epsilon_i \delta_{ij} + \Delta v_{ij} \nonumber \\
    f_{ab} & = \langle \psi_a|\hat{f}|\psi_b \rangle = \epsilon_a \delta_{ab} + \Delta v_{ab} \nonumber \, .
  \end{align}
The second step is to diagonalize the $f_{ij}$ and the $f_{ab}$ block separately. Denoting the 
eigenvector matrices as ${\cal O}$ and ${\cal U}$, one has
 \begin{align}
    \sum_{k} f_{ik}{\cal O}_{kj} = {\cal O}_{ij} \tilde{\epsilon}_j \nonumber \\
    \sum_{c} f_{ac}{\cal U}_{cb} = {\cal U}_{ab} \tilde{\epsilon}_b  \, ,
 \end{align}
where $\tilde{\epsilon}_j$ and $\tilde{\epsilon}_b$ are the eigenvalues of the occupied and unoccupied blocks 
of the Fock matrix, respectively. 
This procedure is known as \emph{semi-canonicalization} in quantum chemistry 
(see e.g. Ref.~\onlinecite{Schweigert/Lotrich/Bartlett:2006}).
The final rSE expression, equivalent to the infinite-order diagrammatic summation, is given by
  \begin{equation}
     E_\text{c}^\text{rSE} = \sum_{ia} \frac{|\tilde{f}_{ia}|^2}{\tilde{\epsilon}_i - \tilde{\epsilon}_a}\, ,
     \label{Eq:E_rSE_final}
  \end{equation}
where $\tilde{f}_{ia}$ correspond to the ``transformed" off-diagonal block of the Fock matrix
 \begin{equation}
    \tilde{f}_{ia} = \sum_{jb}{\cal O^\ast}_{ij} {\cal U^\ast}_{ab} f_{jb} \, .
 \end{equation}
This is a surprisingly simple result: the final rSE expression is formally identical to the 2nd-order
SE one; only that the meaning of the energy eigenvalues and the transition amplitudes has to be
modified. The equivalence of Eq.~(\ref{Eq:E_rSE_final}) to the algebraical expression 
from a direct evaluation of the diagrams in Fig.~\ref{Fig:rSE_diagrams} is demonstrated in appendix~\ref{sec:app/rSE}.

\subsection{\label{sec:th_rPT2}The concept of rPT2 viewed from its diagrammatic representation}

\begin{figure}
     \begin{picture}(200,180)(0,0)
       \put(50,115){\makebox(0,0){\includegraphics[width=0.3\textwidth,clip]{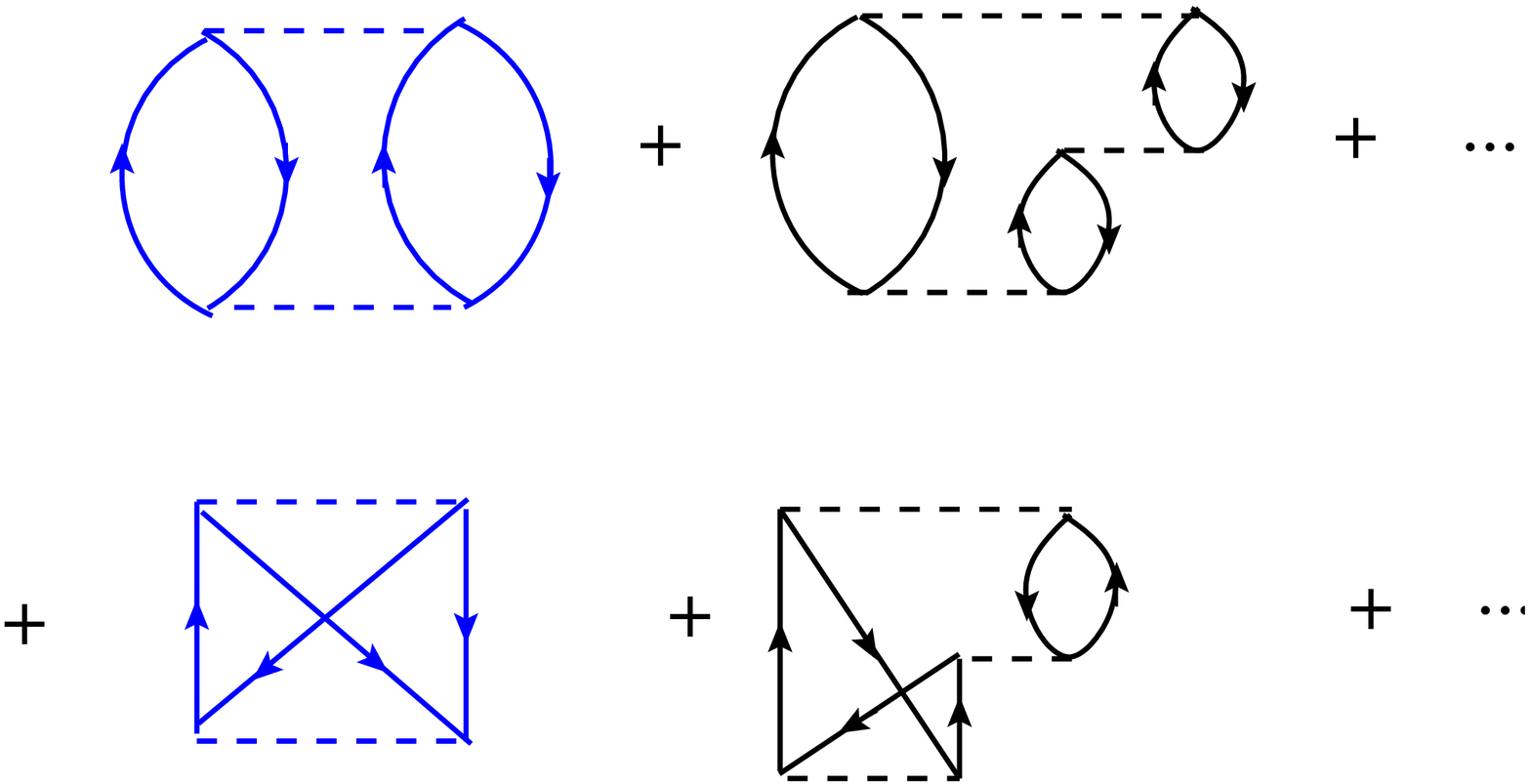}}}
       \put(160,135){ ( = {RPA})}
       \put(160,95){ ( = {SOSEX})}
       \put(75,35){\makebox(0,0){\includegraphics[width=0.4\textwidth,clip]{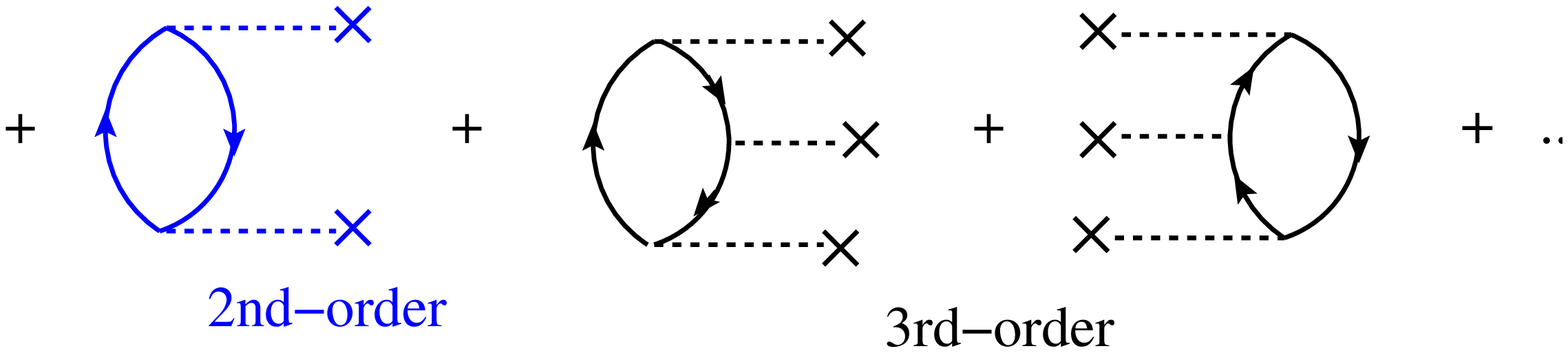}}}
       \put(180,40){ ( = {rSE})}
     \end{picture}
  \caption{(Color online) rPT2 represented in terms of Goldstone diagrams. The three rows of (infinitely summed) diagrams 
          represent the three components of rPT2: RPA, SOSEX, and rSE. The first column shows the (only) three
          terms in normal (bare) 2nd-order Rayleigh-Schr\"odinger perturbation theory based on a KS reference.}
   \label{Fig:rPT2_diagram}
\end{figure}

Initially the RPA+SOSEX and RPA+(r)SE  schemes were developed separately \cite{Grueneis/etal:2009,Ren/etal:2011}  in an effort to improve the accuracy of the RPA method. 
In Ref.~\onlinecite{Paier/etal:2012} it was found that adding both terms to RPA leads to even better
accuracy in general, and that the combined RPA+SOSEX+rSE ($\equiv$rPT2) scheme represents the most balanced 
approach for describing both atomization energies and reaction barrier heights. To elucidate the nature of rPT2, 
the Goldstone diagrams for the three ingredients of this theory are shown together in 
Fig.~\ref{Fig:rPT2_diagram}. All three pieces are characterized by an infinite summation of diagrams with the same
topological structure. The leading terms in the three series are the
second-order direct (Coulomb), the second-order exchange, and the SE term, respectively.
 In other words, these leading terms are exactly the (only) three terms that one would encounter in second-order Rayleigh-Schr\"odinger perturbation theory,
based on an (approximate) KS reference Hamiltonian. Only the SE term would vanish if  the perturbation series were to be build on the HF reference. In essence, the theory is exact at second order,
and for higher-order contributions we follow the strategy of  ``selective summation to infinite order", following 
the spirit of the RPA. This ``infinite-order summation" effectively \emph{renormalizes} the three terms of the (bare) second-order perturbation theory (PT2), represented by the the blue diagrams in Fig.~\ref{Fig:rPT2_diagram}. 
 We expect the renormalized method, i.e. rPT2, to be more generally 
applicable  than the bare PT2, which, e.g., suffers from notorious divergence problems for systems with zero direct gap.\cite{Fetter/Walecka:1971,Grueneis/Marsman/Kresse:2010}

As a perturbation theory, rPT2 will necessarily depend on the reference Hamiltonian, or equivalently a set of
input single-particle orbitals. In practice, rPT2 works best when based on Kohn-Sham Hamiltonians, that yield a smaller
gap than generalized KS or HF ones.  This is directly related to the fact that the underbinding error
of RPA will be even more pronounced for HF or generalized KS reference Hamiltonians, as evidenced by the
significant RPA@HF error for the G2 atomization energies, \cite{Ren/etal:2012} and the severely underestimated
RPA@HF ($40\%$) C$_6$ coefficients \cite{Ren/unpublished} (here and in the following, we use 
``\emph{method@reference}" to denote which \emph{method} is based on which \emph{reference} state).
For a variety of KS Hamiltonians (i.e. with local, multiplicative potentials),
RPA results were found to be insensitive to the actual choice of the reference Hamiltonian.
\cite{Harl/Schimka/Kresse:2010,Eshuis/Furche:2011} In this work,  we will therefore choose the most popular 
non-empirical GGA functional PBE as the reference; also to be consistent with our previous work.
\cite{Ren/etal:2011,Paier/etal:2012,Ren/etal:2012b} The insensitivity of RPA to reference KS Hamiltonians
carries over to rPT2.

\section{\label{sec:results}Results}
In this section we will benchmark the performance of rPT2 for weak interaction energies (rare-gas dimers,
S22 and S66 test sets by Hobza and coworkers\cite{Jurecka/etal:2006,Rezac/etal:2011}), atomization energies 
(from the G2-I test set by Curtiss \textit{et al.}\cite{Curtiss/etal:1997,Curtiss/etal:2005}), and 
chemical reaction barrier heights (38 hydrogen-transfer and 38 non-hydrogen-transfer chemical reactions of 
Truhlar and coworkers \cite{Zhao/Nuria/Truhlar:2005,Zhao/Truhlar:2006}). All calculations were performed with 
the local-orbital based all-electron FHI-aims code.\cite{Blum/etal:2009,Ren/etal:2012}  
As mentioned in 
section~\ref{sec:th_rpa+sosex}, the SOSEX term in this work corresponds to ``AC-SOSEX" based on Eq.~(\ref{Eq:EcSOSEX_simpl}).  For brevity we will simply refer to it as SOSEX in the following.
For the frequency integration in our RPA and SOSEX calculations, we use a modified Gauss-Legendre grid 
\cite{Ren/etal:2012} with 40 points. For the $\lambda$ integration in Eq.~(\ref{Eq:W_bar}), we use a normal
Gauss-Legendre grid with 5 points. These settings guarantee sufficient accuracy for the benchmark studies
presented in this work.  The basis 
sets employed in the calculations will be specified later when discussing the results. Convergence tests are shown in Appendix~\ref{sec:app/basis}.

\subsection{\label{sec:vdW}Weak interactions}

One prominent feature of RPA-based approaches is that the ubiquitous vdW interactions can be captured
in a seamless manner. \cite{Szabo/Ostlund:1977,Dobson:1994}  The long-range behavior of the RPA interaction
energy between two closed-shell molecular systems decays as $C_6/R^6$ where the $C_6$ value is dictated by 
the RPA polarizability of the monomer. \cite{Dobson:1994,Dobson/Could:2012} Many-body terms that go beyond
the pair-wise summation are also automatically contained in this approach. \cite{Lu/Nguyen/Galli:2010} 

Benchmarking the performance of RPA and related methods for vdW bonded systems has been a very active 
enterprise. 
\cite{Furche/Voorhis:2005,Harl/Kresse:2008,Janesko/Henderson/Scuseria:2009,Toulouse/etal:2009,Lu/Li/Rocca/Galli:2009,Li/Lu/Nguyen/Galli:2010,Zhu/etal:2010,Jansen/etal:2010,Toulouse/etal:2011,Ren/etal:2011,Eshuis/Furche:2011} 
It has been demonstrated that the standard RPA approach 
exhibits a systematic underbinding behavior for molecules, in particular vdW bonded ones. 
\cite{Ren/etal:2011}  We have previously shown that SE-type corrections ameliorate this problem,
\cite{Ren/etal:2011} but the influence of the SOSEX correction has not been systematically 
benchmarked for vdW systems yet, with the exception of He$_2$ and Ne$_2$.\cite{Paier/etal:2010}
It is therefore interesting and timely to examine how rPT2, that combines both types of corrections, performs for 
noncovalent interactions. Some rPT2 results for Ar$_2$ and S22 have featured in our recent review on RPA. \cite{Ren/etal:2012b} Here we extend the benchmark study to other rare-gas  dimers and also the larger S66 test set.

\subsubsection{Rare-gas dimers}
\label{sec:rare-gas-dimer}
\begin{figure}
\includegraphics[width=0.45\textwidth]{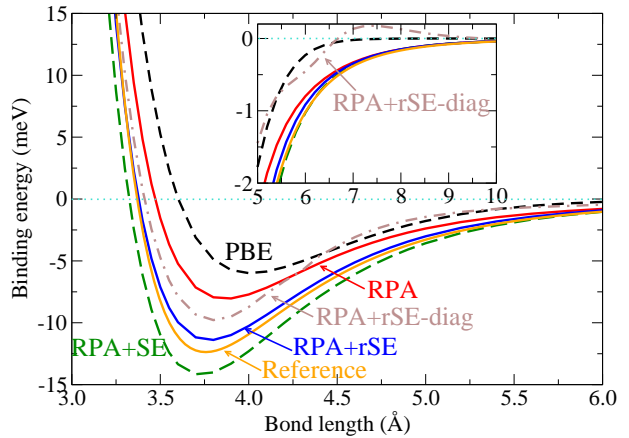}
\caption{(Color online) 
Binding energy curves for Ar$_2$ computed with PBE and RPA-based approaches (standard RPA, RPA+SE, 
RPA+rSE, and RPA+rSE-diag based on PBE), in comparison with the Tang-Toennies reference curve. The results are obtained using the Gaussian ``aug-cc-pV6Z" \cite{Dunning:1989} basis set. The basis set superposition error (BSSE) is corrected here and in all following calculations using the counterpoise correction scheme.\cite{Boys/Bernardi:1970} }
\label{Fig:Ar2_rSE-d}
\end{figure}

First, we demonstrate the pathological behavior of rSE-diag for weak interactions, highlighting the importance of including the ``off-diagonal" terms in the rSE summation to make the theory invariant with respect to orbital rotations. 
In Fig.~\ref{Fig:Ar2_rSE-d} the binding energy of Ar$_2$ is plotted for PBE, RPA, and RPA plus different versions of single excitation corrections (RPA+SE, RPA+rSE-diag, RPA+rSE). While PBE, RPA, and RPA+SE all show their 
characteristic behaviors, the behavior of RPA+rSE-diag is weird. The binding energy curve develops unphysical undulations away from equilibrium. Moreover, the
asymptotic limit does not follow the correct $1/R^6$ behavior, and the curve even reaches above the energy zero
at large bonding distances (see the inset of Fig.~\ref{Fig:Ar2_rSE-d}). Naturally, this problem also carries 
over to rPT2-diag (not shown). It is reassuring, however, to
observe that this pathological behavior disappears in the upgraded RPA+rSE scheme, which yields a binding energy 
curve in close agreement with the Tang-Toennies reference curve,\cite{Tang/Toennies:2003} obtained
from a simple analytical model with
experimental equilibrium bond distance and binding energy as input parameters. This model can accurately
reproduce empirical data \cite{Tang/Toennies:2003} and agrees excellently with high-level quantum-chemical,
e.g., CCSD(T) calculations.  \cite{Klopper/Noga:1995,Laschuk/etal:2003}
Coming back to the rSE discussion, the pathological behavior
is thus caused by the diagonal approximation, and not inherent to the rSE scheme itself. 
In the remainder of our discussion on weakly interacting systems, we therefore only present results for the upgraded RPA+rSE and rPT2 schemes.

\begin{figure*}
\includegraphics[width=0.6\textwidth,clip]{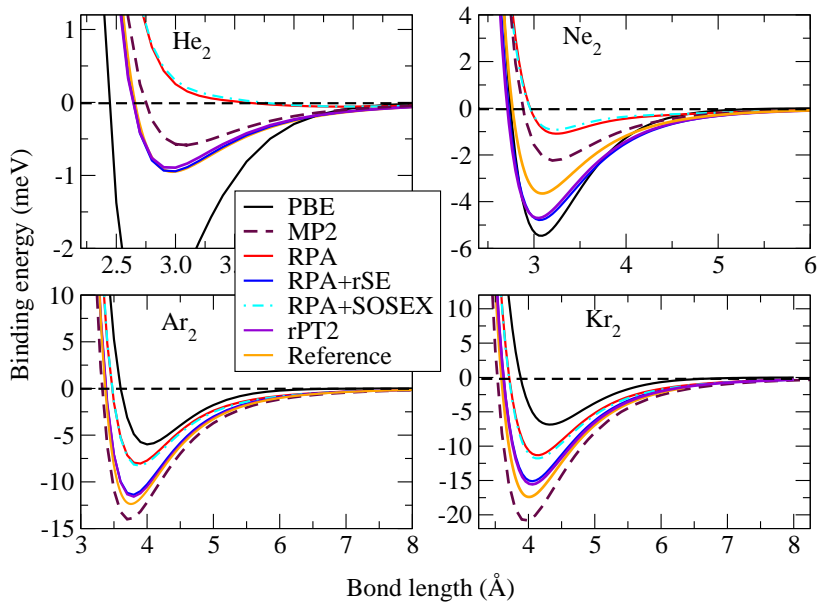}
\caption{(Color online) Binding energy curves for rare-gas dimers computed with RPA-based approaches, in comparison with
    PBE, MP2, and the Tang-Toennies reference curves. He$_2$, Ne$_2$, and Ar$_2$ results are obtained using 
    the aug-cc-pV6Z basis set, and Kr$_2$ using the aug-cc-pV5Z basis set. All RPA-type calculations are based 
    on the PBE reference.}
  \label{Fig:rare_gas_dimer}
\end{figure*}

The full set of binding energy curves for He$_2$, Ne$_2$, Ar$_2$, and Kr$_2$ obtained with PBE, MP2, RPA, rPT2, 
as well as the ``intermediate" schemes RPA+rSE and RPA+SOSEX are then shown in Fig.~\ref{Fig:rare_gas_dimer}.
PBE does not contain long-range dispersion interactions by construction, and therefore decays too fast at large
separations. Around the equilibrium region, PBE vastly overbinds He$_2$ and Ne$_2$, and underbinds Ar$_2$ and 
Kr$_2$. MP2 shows the opposite trend, although it performs better at a quantitative level. RPA systematically 
underbinds all dimers. This underbinding is most significant for He$_2$ and Ne$_2$. 
Adding the rSE correction leads to a substantial improvement for all dimers. With the largest available 
Dunning Gaussian basis sets \cite{Dunning:1989} (aug-cc-pV6Z for He, Ne, Ar and aug-cc-pV5Z for Kr), 
RPA+rSE shows nearly perfect agreement with the reference curve for He$_2$, overshoots a little bit for Ne$_2$, and 
slightly underbinds Ar$_2$ and Kr$_2$. The SOSEX correction, on the other hand,  has very little effect on the 
binding energies of these purely dispersion-bonded systems. As a result, rPT2 lies almost
on top of RPA+rSE.  The overall accuracy of RPA+rSE and rPT2 for rare-gas dimers is very satisfactory, 
in particular since no adjustable parameters are used in these schemes. 

\subsubsection{S22 and S66 test sets}

A widely used benchmark set for weak interactions are the S22 molecular complexes designed by 
Jure\v{c}ka et al., \cite{Jurecka/etal:2006} for which accurate reference interaction energies obtained
using the the CCSD(T) method are 
available. \cite{Takatani/etal:2010} This molecular test set includes the most common types of non-covalent 
interactions: hydrogen bonding, dispersion-dominated bonding, and those of mixed character.  The performance of RPA
and some of the RPA-related methods have been benchmarked for this test set. 
\cite{Zhu/etal:2010,Eshuis/Furche:2011,Ren/etal:2011,Eshuis/Furche:2012}
Similar to correlated quantum chemical methods, the quality 
of basis sets for RPA calculations is a significant issue.\cite{Furche:2001,Eshuis/Furche:2012,Fabinao/DellaSala:2012}
Using our numerical atomic orbital (NAO) \textit{tier} 4 basis plus additional diffuse Gaussian functions from  the
aug-cc-pV5Z set (denoted as ``\textit{tier} 4 + a5Z-d"\cite{Ren/etal:2012}; see also Appendix~\ref{sec:app/basis}), we obtained a 
mean absolute error (MAE) of 0.90 kcal/mol in RPA@PBE 
for S22, fairly close to the 0.79 kcal/mol reported by Eshuis and Furche \cite{Eshuis/Furche:2012} 
using Dunning's Gaussian basis sets extrapolated to the complete
basis set (CBS) limit. In Appendix~\ref{sec:app/basis} the convergence behavior of these two types of basis
sets is shown for the methane dimer.
In this work we will continue to use the ``\textit{tier} 4 + a5Z-d" basis set, bearing in mind that the absolute numbers could carry an uncertainty of 0.1 kcal/mol (4 meV), which  will however not affect our discussion here.

\begin{figure}
\includegraphics[width=0.48\textwidth,clip]{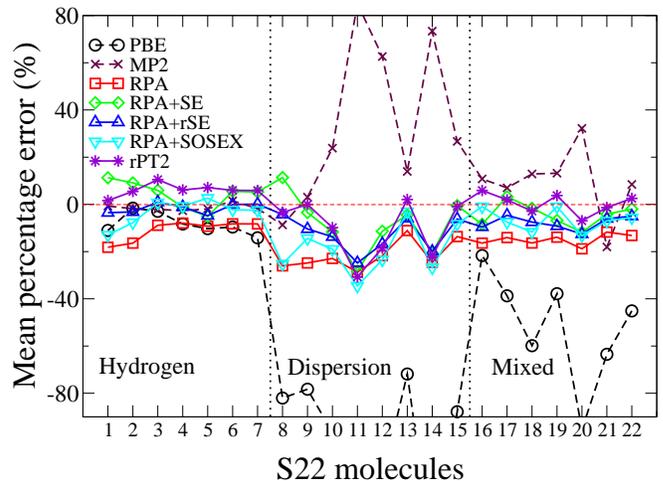}
\caption{(Color online) The percentage errors for the S22 test set for RPA-derived computational schemes 
(based on  PBE 
reference orbitals), in comparison to PBE and MP2. 
The CCSD(T)/CBS
results of Takatani {\it et al.} \cite{Takatani/etal:2010} are used as reference. Lines are guides to the eye.}
\label{Fig:S22_PE}
\end{figure}

In Fig.~\ref{Fig:S22_PE} the relative errors from RPA+rSE, RPA+SOSEX, and rPT2
are presented for each individual molecule of the S22 set. Results from RPA and RPA+SE, as well as from
PBE and MP2 are also included for comparison. PBE and MP2 are both performing
well for hydrogen-bonded molecules where the electrostatic interactions dominate, but PBE underbinds 
the dispersion-dominated and those of
mixed-character significantly, while the opposite is true for MP2. RPA-based methods are performing much better than PBE and MP2 for these two types of interactions. RPA+rSE falls between RPA and RPA+SE, although it lies
closer to RPA+SE. For hydrogen-bonded molecules, RPA+rSE improves 
over RPA+SE, with the latter overbinding these molecules noticeably. Moreover, it
is interesting to note that RPA+SOSEX improves over RPA appreciably for hydrogen-   and mixed-bonding, but 
much less so for dispersion-bonded molecules. This is consistent with its performance for rare-gas
dimers. Now, combining rSE and SOSEX, rPT2 performs equally well or better for dispersion-dominated and mixed-bonding, but overshoots significantly for hydrogen-bonding. So far this is the only case we have found, for which  
combining rSE and SOSEX worsens the description.
Finally we note that for $\pi$-stacked systems like the benzene dimer (\# 11) RPA gives a substantial error, 
but neither rSE nor SOSEX significantly improves upon RPA. This warrants further attention in future studies.

\begin{figure}
\includegraphics[width=0.45\textwidth,clip]{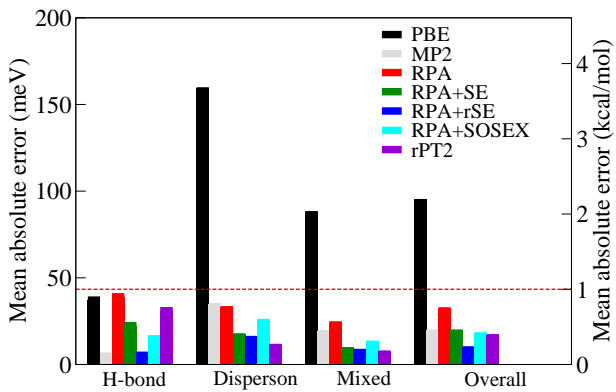}
\caption{
(Color online) MAEs (in both meV and kcal/mol) for the S66 test set given by RPA, rPT2 and related schemes (based on  PBE  reference orbitals), in addition to PBE and MP2. The CCSD(T) results of Rezac {\it et al.} \cite{Rezac/etal:2011} at the CBS limit are used here as reference.}
\label{Fig:S66_mae}
\end{figure}

Recently the S22 test set has been extended to an even larger, more comprehensive and balanced test set 
called S66.\cite{Rezac/etal:2011}  This overcomes several shortcomings of S22, e.g. the strong bias towards 
nucleic-acid-like structures.
We also performed benchmark calculations with RPA, rPT2, and related computational schemes for this test set,
and the results are presented in Fig.~\ref{Fig:S66_mae}. The overall performance for S66 is very similar to that observed for S22. In brief, RPA+rSE
performs better (or slightly better) than RPA+SE, which itself is a
significant improvement over the standard RPA method. Adding SOSEX, the resultant rPT2
approach performs even (slightly) better than RPA+rSE for dispersion and mixed interactions. However,
this is not the case for hydrogen bonds, where rPT2 clearly overshoots and the
strength of hydrogen bonds becomes overestimated. Overall, for weak interactions RPA+rSE outperforms other 
computational schemes benchmarked here, and yields a MAE of 10.1 meV (or 0.23 kcal/mol).
%

\subsection{G2 atomization energies}
\begin{figure}
\includegraphics[width=0.48\textwidth,clip]{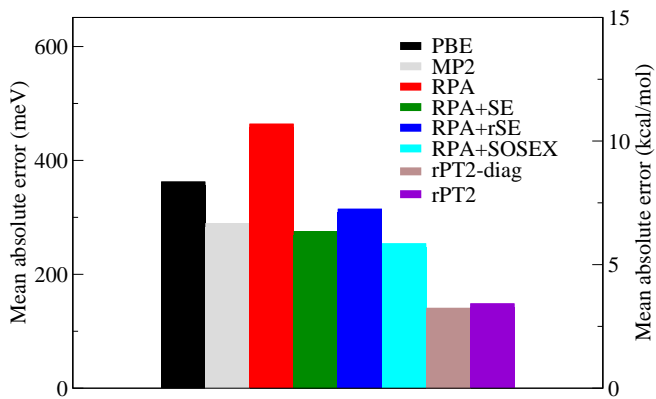}
\caption{The MAEs (in both meV and kcal/mol) of the G2-I atomization energies \cite{Curtiss/etal:1997} obtained with 
PBE, MP2, RPA, rPT2, and related methods. The Gaussian cc-pV6Z basis set \cite{Dunning:1989} was used in all calculations. Reference data are from Ref.~\onlinecite{Feller/Peterson:1998}. }
\label{Fig:G2-I}
\end{figure}

The atomization energy of molecules is a key quantity in thermochemistry. RPA has been tested for this quantity
in early works,\cite{Furche:2001,Paier/etal:2010} where a pronounced underbinding behavior was observed. In a recent
work, Paier \textit{et al.} \cite{Paier/etal:2012} reported a detailed study of the atomization energies of the 
G2-I set\cite{Curtiss/etal:1997} using RPA and its variants, including the rPT2-diag scheme as 
discussed before. To test the influence of the off-diagonal elements of rSE in the rPT2 scheme,  we present in Fig.~\ref{Fig:G2-I} the MAEs for  RPA, rPT2-diag, rPT2, and related methods. Some of these results were already included in our recent review paper on RPA.\cite{Ren/etal:2012b}  
In brief, the MAE for RPA is significantly reduced when adding the (r)SE 
or SOSEX corrections. In this case, RPA+rSE yields a slightly larger MAE than RPA+SE. Combining the rSE and SOSEX
corrections, rPT2 reduces the MAE further by a factor of two. In contrast to the nonbonded  interactions discussed in the previous section, the difference between rPT2 and rPT2-diag is small (0.18 kcal/mol or 8 meV difference in
MAE). 
validating our previous conclusions regarding the atomization energies in Ref.~\onlinecite{Paier/etal:2012} that were based on the rPT2-diag scheme. 

In this context we would like to warn that, despite the success of RPA+SOSEX and rPT2 for describing 
the atomization energies on average, adding SOSEX to RPA makes things worse (more underbinding) 
for certain molecules (in particular O$_2$ and N$_2$), and this problem also carries over to rPT2. 
A detailed investigation of this issue is beyond the scope of this paper, and will be carried out
in future work.

\subsection{Barrier heights}
\begin{figure}
\includegraphics[width=0.48\textwidth,clip]{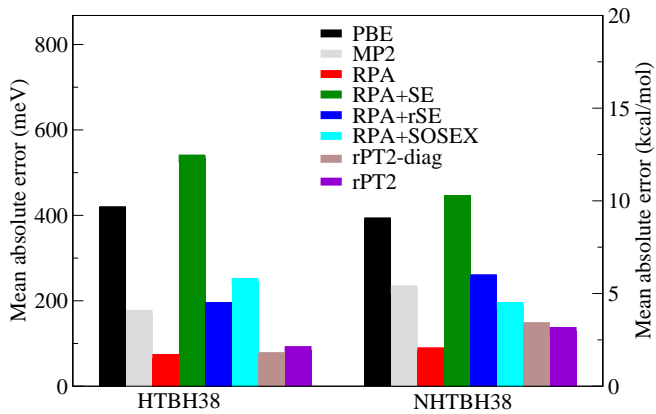}
\caption{
(Color online) The MAEs (in both meV and kcal/mol) of the HTBH38 and NHTBH38 test sets for barrier heights, 
obtained with PBE, MP2, RPA, rPT2, and related methods (based on PBE). Reference data are from 
Ref.~\onlinecite{Zhao/Nuria/Truhlar:2005,Zhao/Truhlar:2006}. Gaussian cc-pV6Z basis sets 
were used in all calculations. }
\label{Fig:BH76}
\end{figure}

To complete  our discussion, we address chemical reaction barrier heights. For this purpose we chose
the HTBH38 and NHTBH38 test set of Truhlar and coworkers.
\cite{Zhao/Nuria/Truhlar:2005,Zhao/Truhlar:2006} RPA-based methods were benchmarked in previous
studies\cite{Paier/etal:2012,Eshuis/Bates/Furche:2012,Ren/etal:2012b}, and we here revisit this set
with the upgraded version of rPT2. The MAEs for our different schemes are shown 
in Fig.~\ref{Fig:BH76}. Standard RPA performs remarkably well for reaction barrier
heights compared to all alternatives. This has been rationalized by Henderson and 
Scuseria \cite{Henderson/Scuseria:2010} to be due to the inherent self-correlation error in RPA that mimics ``static correlation" (i.e. the (near) degeneracy of two (or more) determinants), leading to an excellent description of the transition states due to partial error cancellation.
Unfortunately, any attempt to correct RPA  deteriorates its performance in this case. In particular, the RPA+SE method 
provides a bad description of the transition states, resulting in errors that are even larger than in PBE. The RPA+SE 
error reduces when the SE term is renormalized in RPA+rSE. The errors in RPA+rSE and RPA+SOSEX tend 
to cancel each other, and by combining the two schemes, rPT2 gives a much more satisfactory 
description of the barrier heights. 
Similar to the G2-I test set, the difference between rPT2 and rPT2-diag is small (0.33 kcal/mol for HTBH38 
and 0.25 kcal/mol for NHTBH38 in MAE) compared to the variation among other schemes.

\section{\label{sec:conc}Conclusions}
In summary, the rPT2 method comprises an infinite summation of three distinct series of diagrams:  RPA, SOSEX, and rSE.  As is obvious from its diagrammatic representation, rPT2 can be viewed as a \emph{renormalization} of bare second-order perturbation theory -- the latter being the leading term of rPT2. In this work we derived an alternative way
to express the SOSEX correlation energy, discussed in detail how to sum up the ``off-diagonal" elements
in rSE, which were neglected in previous works, and presented the concept of rPT2 from a diagrammatic point
of view. We benchmarked the performance of rPT2 and related approaches (RPA+rSE, RPA+SOSEX), focusing on weakly interacting molecules. We found that rPT2 works well for dispersion and mixed-type interactions, but for hydrogen bonds it over-corrects the underbinding behavior of RPA. We also examined the influence of the previously neglected ``off-diagonal" elements in the rSE  correction and found that, for weak interactions, it is crucial to include them, whereas for atomization 
energy and reaction barrier heights, the off-diagonal elements only have a minor effect. We also found that the SOSEX correction improves the description of electrostatic interactions substantially, but has very little effect on dispersion interactions. rSE, on the other hand, leads to a better description of both electrostatic and dispersion interactions. 

Overall rPT2 provides a conceptually appealing, and diagrammatically systematic way for going
beyond RPA. Although it does not always deliver the best accuracy in every single case compared to other RPA-based
approaches, it provides the most ``balanced" description across various different
electronic and chemical environments. We thus consider the rPT2 scheme as a natural step for extending 
and improving the RPA method. The successes and shortcomings of rPT2 documented in this work provide
a useful basis for developing more accurate, robust, and generally applicable electronic structure methods
in the coming years.

\section*{ACKNOWLEDGMENTS}
We thank Joachim Paier for making available his SOSEX numbers generated using \GDV, and 
Jonathan E. Moussa for a critical reading of the manuscript and pointing out to us the distinction 
between SOSEX and AC-SOSEX.  The work at Rice University was supported by the US Department
of Energy, Office of Basic Energy Sciences (Grant No. DEFG02-09ER16053)
and the Welch Foundation (Grant No. C-0036).

\begin{appendix}
\section{\label{sec:appendix}Implementation of AC-SOSEX in FHI-aims}

The RPA implementation in FHI-aims has been described in detail in Ref.~\onlinecite{Ren/etal:2012}. 
Here we will give a brief account of the SOSEX implementation in our code. The energy expression 
that we would like to evaluate is
\begin{align}
   E_c^\text{AC-SOSEX} = -\frac{1}{2\pi} \int_0^\infty d\omega & \sum_{ia,jb} \langle ij|ba \rangle 
       \langle ij |\bar{W}(i\omega)| ab\rangle \times \nonumber \\
        &  {\cal F }_{ia}(i\omega) {\cal F }_{jb}(i\omega) \,
   \label{Eq:app_E_SOSEX}
\end{align}
where  $\langle ij|ba \rangle$ are the two-electron Coulomb
integrals defined in Eq.~(\ref{Eq:2eri}), and $\langle ij |\bar{W}(i\omega)| ab\rangle$ are the
corresponding (coupling-constant-averaged) screened Coulomb integrals.
The frequency-dependent factor ${\cal F }_{ia}(i\omega)$ is defined in Eq.~(\ref{Eq:factor}).

In analogy to the RPA case, the basic technique  to evaluate the two-electron integrals
in our code is the resolution-of-identity. We chose the Coulomb metric, denoted ``RI-V" in the following.  
Here we would like to emphasize that ``RI-V" is a highly accurate method, and the error incurred
thereby is vanishingly small for practical purposes (see Ref.~\onlinecite{Ren/etal:2012} for detail benchmarks).
In RI-V, the bare two-electron integrals are computed as
 \begin{equation}
    \label{Eq:RI_Vmatr}
    \langle ij|ab\rangle = \sum_{\mu\nu} (ia|\mu) V_{\mu\nu}^{-1} (\nu|ib) \, \\
 \end{equation}
where 
 \begin{equation}
   (ia|\mu) = \iint d\bfr d\bfrp \frac{\psi_i(\bfr)\psi_a(\bfr)P_\mu(\bfrp)}{|\bfr-\bfrp|} \, ,
  \label{Eq:RI_V_3index}
 \end{equation}
and 
 \begin{equation}
   V_{\mu\nu} = \iint d\bfr d\bfrp \frac{P_\mu(\bfr)P_\nu(\bfrp)}{|\bfr-\bfrp|} \, .
 \end{equation}
Here $\psi_p$ are canonical single-particle spin-orbitals, and $P_\mu(\bfr)$ are a set of
suitably constructed auxiliary basis functions.\cite{Ren/etal:2012}  For notational simplicity all orbitals are assumed
to be real.

In practice, we decompose the $V^{-1}$ matrix in Eq.~(\ref{Eq:RI_Vmatr}) into
the product of its square roots, and combine each three-index integral with a square root. This gives
 \begin{equation}
    \langle ij|ab\rangle = \sum_{\mu} O_{ia}^\mu O_{jb}^\mu \, \\
    \label{Eq:RI_Vmatr1}
 \end{equation}
with
 \begin{equation}
    O_{ia}^\mu = \sum_{\nu} (ia|\nu) V^{-1/2}_{\nu \mu}\, .
  \label{Eq:RI_V_coeff}
 \end{equation}

As discussed in the context of the $GW$ implementation in FHI-aims, \cite{Ren/etal:2012}  the ``RI-V" technique can 
be used to treat the screened two-electron Coulomb integrals as well.  In this case we have
\begin{equation}
\langle ij |\bar{W}(i\omega)|ab\rangle = \sum_{\mu,\nu} O_{ia}^\mu \bar{\cal E}^{-1}_{\mu\nu}(i\omega)O_{jb}^\mu 
\label{Eq:W_matr_dielec}
\end{equation}
where $\bar{\cal E}$ is the coupling-constant averaged dielectric functions, formally linked to
the screened Coulomb matrix by 
 \begin{equation}
  \bar{\cal E}^{-1}(i\omega)=V^{-1/2}\bar{W}(i\omega)V^{-1/2}\, .
   \label{Eq:dielec_func}
 \end{equation}
In Eq.~(\ref{Eq:dielec_func}), $\bar{W}(i\omega)$ is the screened Coulomb interaction matrix represented in terms of 
the auxiliary basis set,
  \begin{equation}
      \bar{W}_{\mu\nu}(i\omega) = \iint d\bfr d\bfrp P_\mu(\bfr) \bar{W}_{\mu\nu}(\bfr,\bfrp,i\omega) P_\nu(\bfrp).
  \end{equation}

For convenience, we introduce a quantity $\Pi(i\omega)=v^{1/2}\chi_0(i\omega)v^{1/2}$, where
$\chi_0(i\omega)$ is the independent density response function defined in Eq.~(\ref{Eq:chi_0_realspace}). Using Eqs.~(\ref{Eq:chi_0_realspace}), (\ref{Eq:RI_V_3index}), and (\ref{Eq:RI_V_coeff}),
one can easily obtain the matrix representation of $\Pi(i\omega)$ in the auxiliary basis 
  \begin{equation}
        \Pi_{\mu\nu}(i\omega) = \sum_{ia}\frac{2(\epsilon_i-\epsilon_a)}
       {\omega^2+(\epsilon_i-\epsilon_a)^2}O_{ia}^\mu O_{ia}^\nu \,
  \end{equation}
where $\epsilon_i$ and $\epsilon_a$ are occupied and unoccupied single-particle orbital energies, respectively.
Using Eq.~(\ref{Eq:W_lambda}), the matrix form of $\bar{\cal E}^{-1}$ becomes
  \begin{equation}
     \bar{\cal E}^{-1}(i\omega) = \int_0^1 d\lambda \left[1-\lambda \Pi(i\omega)\right]^{-1}\lambda  \, .
     \label{Eq:EPS_integral}
  \end{equation}
The $\lambda$-integration in Eq.~(\ref{Eq:EPS_integral}) can be accurately computed using a Gauss-Legendre 
quadrature with 5-6 grid points.

Combining Eqs~(\ref{Eq:app_E_SOSEX}), (\ref{Eq:RI_Vmatr1}), and (\ref{Eq:W_matr_dielec}), the final expression
for the RI-SOSEX energy is
 \begin{widetext}
   \begin{eqnarray}
     E_\text{c}^\text{SOSEX} &=& -\frac{1}{2\pi} \int_{0}^\infty d\omega \sum_{ij,ab} 
    \left[ \left(\sum_{\mu}O_{ia}^\mu O_{jb}^\mu\right)
    \left(\sum_{\nu\gamma}O_{ia}^\nu \bar{\cal E}_{\nu\gamma}^{-1}(i\omega) O_{jb}^{\gamma}\right)\right]
     {\cal F}_{ia}(i\omega) {\cal F}_{jb}(i\omega)\, .
    \label{Eq:SOSEX_RI}
  \end{eqnarray}
 \end{widetext}
The computational effort for evaluating Eq.~(\ref{Eq:SOSEX_RI}) formally scales as $O(N^5)$, where $N$ is the system
size.

\section{\label{app:SE_deriv} Derivation of the single excitation contribution to the 2nd-order correlation energy}
In this section we derive Eq.~(\ref{Eq:SE}) that is presented in the main part of this paper -- the single excitation contribution to the 2nd-order correlation energy -- from Rayleigh-Schr\"odinger perturbation theory (RSPT).   The interacting $N$-electron system at hand is governed  by the Hamiltonian 
 \begin{equation}
   \hat{H} = \sum_{j=1}^{N}\left[-\frac{1}{2}\nabla^2_j + \hat{v}_\text{ext} (\bfr_j) \right]  + 
     \sum_{j<k}^{N} \frac{1}{|\bfr_j -\bfr_k|}, \nonumber
 \end{equation}
where  $\hat{v}_\text{ext}(\bfr)$ is a local, multiplicative external potential. In RSPT, $\hat{H}$ is partitioned into a non-interacting mean-field  Hamiltonian $\hat{H}^0$ and an interacting perturbation $\hat{H}'$,
 \begin{eqnarray} 
   \hat{H} & =& \hat{H}^0 + \hat{H}'    \nonumber  \\
   \hat{H}^0 &=& \sum_{j=1}^{N} \hat{h}^0 (j) = \sum_{j=1}^{N} \left[-\frac{1}{2}\nabla^2_j + \hat{v}_\text{ext} (\bfr_j) + 
         \hat{v}^\text{MF}_j \right] \nonumber \\
   \hat{H}' &=& \sum_{j<k}^{N} \frac{1}{|\bfr_j -\bfr_k|} - \sum_{j=1}^N \hat{v}^\text{MF}_j. \nonumber
 \end{eqnarray}
Here $\hat{v}^\text{MF}$ is any mean-field potential, which can be non-local, as 
in the case of Hartree-Fock (HF) theory, or local, as in the case of Kohn-Sham (KS) theory.

Suppose the solution of the single-particle Hamiltonian $\hat{h}^0$ is known 
  \begin{equation}
    \hat{h}^0 |\psi_p\rangle = \epsilon_p |\psi_p\rangle,
   \label{Eq:SP_eigeneq}
  \end{equation}
then the solution of the non-interacting many-body Hamiltonian $H^0$ follows directly
 \begin{equation}
   \hat{H}^0 |\Phi_n \rangle = E^{(0)}_n |\Phi_n \rangle  \nonumber .
 \end{equation}
The $|\Phi_n \rangle$ are single Slater determinants formed from $N$ of the spin orbitals $|p\rangle = |\psi_p\rangle$ determined in Eq.~(\ref{Eq:SP_eigeneq}).  These Slater determinants can be distinguished according to their excitation level: the ground-state configuration $|\Phi_0\rangle$, singly excited  configurations $|\Phi_i^a\rangle$, doubly excited configurations $|\Phi_{ij}^{ab}\rangle$, etc., where $i,j,k, \dots$ denotes occupied orbitals and $a,b,c,\dots$ unoccupied ones. Following standard perturbation theory, the  single-excitation (SE) contribution to the 2nd-order correlation energy is given by
 \begin{eqnarray}
  E^\text{SE}_c& =& \sum_{i}\sum_a\frac{|\langle\Phi_0|\hat{H}'|\Phi_i^a\rangle|^2}{E^{(0)}_0 - E^{(0)}_{ia}} \nonumber \\
               & =&\sum_{i}\sum_a\frac{|\langle\Phi_0|\sum_{j<k}^{N} \frac{1}{|\bfr_j -\bfr_k|} - \sum_{j=1}^N \hat{v}^\text{MF}_j|\Phi_i^a\rangle|^2}{\epsilon_i - \epsilon_a} \nonumber \\
  \label{Eq:SE_expression}
 \end{eqnarray}
where we have used the fact $E^{(0)}_0 - E^{(0)}_{i,a} = \epsilon_i - \epsilon_a$.

To proceed, the numerator of Eq.~(\ref{Eq:SE_expression}) needs to be evaluated. This can most easily be done using  second-quantization
  \begin{eqnarray}
    \sum_{j<k}^{N} \frac{1}{|\bfr_j -\bfr_k|} & \rightarrow & \frac{1}{2}\sum_{pqrs} \langle pq|rs \rangle c_p^{\dagger} 
             c_q^{\dagger}c_sc_r, \nonumber \\
    \sum_{i=j}^N \hat{v}^\text{MF}_j & \rightarrow & \sum_{pq} \langle p|\hat{v}^\text{MF}|q \rangle c_p^{\dagger} c_q, \nonumber
  \end{eqnarray}
where $p,q,r,s$ are arbitrary spin-orbitals from Eq.~(\ref{Eq:SP_eigeneq}), $c_p^{\dagger}$ and $c_q$, etc.
are the electron creation and annihilation operators, and $ \langle pq|rs \rangle$ the two-electron Coulomb integrals
  \begin{equation}
      \langle pq|rs \rangle = \int d\bfr d\bfrp \frac{\psi_p^\ast(\bfr)\psi_r(\bfr)\psi_q^\ast(\bfrp)\psi_s(\bfrp)}
            {|\bfr-\bfrp|}. \nonumber
  \end{equation}
The expectation value of the two-particle Coulomb operator between the ground-state configuration $\Phi_0$ and the single excitation $\Phi_i^a$ is given by
  \begin{eqnarray}
    \displaystyle
   \langle\Phi_0| \frac{1}{2}\sum_{pqrs} \langle pq|rs \rangle c_p^{\dagger}c_q^{\dagger}c_sc_r |\Phi_i^a\rangle & =&
    \sum_p^\text{occ} \left[ \langle ip|ap\rangle - \langle ip|pa\rangle \right] \nonumber \\
      & =& \langle \psi_i |\hat{v}^\text{HF}|\psi_a\rangle
   \label{Eq:TP_operator}
 \end{eqnarray}
where $v^\text{HF}$ is the HF single-particle potential. 

The expectation value of the mean-field single-particle operator $\hat{v}^\text{MF}$, on the other hand, is given by
 \begin{equation}
    \langle\Phi_0|  \sum_{pq} \langle p|\hat{v}^\text{MF}|q \rangle c_p^{\dagger} c_q | \Phi_i^a\rangle 
    = \langle \psi_i |\hat{v}^\text{MF}|\psi_a\rangle 
   \label{Eq:SP_operator}
 \end{equation}
Combining Eqs.~(\ref{Eq:SE_expression}), (\ref{Eq:TP_operator}), and (\ref{Eq:SP_operator}), one gets
\begin{eqnarray}
  E_{c}^\text{SE} &=& 
   \sum_{i}\sum_a
           \frac{|\langle \psi_i |\hat{v}^\text{HF}-\hat{v}^\text{MF}|\psi_a\rangle|^2}{\epsilon_i - \epsilon_a} 
          \nonumber \\
   &=& \sum_{i}\sum_a \frac{|\Delta v_{ia}|^2}{\epsilon_i - \epsilon_a},
  \label{Eq:2nd_cSE}
\end{eqnarray}
where $\Delta v_{ia}$ is the matrix element of the difference between the HF potential $\hat{v}^\text{HF}$ and 
the single-particle mean-field potential $\hat{v}^\text{MF}$ in question.

Observing that the $\psi$'s are eigenstates of $\hat{h}^0=-\frac{1}{2}\nabla^2 + v_\text{ext}+v^\text{MF}$, and hence all non-diagonal elements $\langle\psi_i|\hat{h}^0|\psi_a\rangle$ are zero, one can alternatively express 
Eq.~(\ref{Eq:2nd_cSE}) as
\begin{eqnarray}
  E_{c}^\text{SE} &=& 
    \sum_{i}\sum_a\frac{|\langle \psi_i |-\frac{1}{2}\nabla^2 + \hat{v}_\text{ext} + 
       \hat{v}^\text{HF} |\psi_a\rangle|^2}{\epsilon_i - \epsilon_a} \nonumber \\
  &=& \sum_{i}\sum_a
           \frac{|\langle \psi_i |\hat{f}|\psi_a\rangle|^2}{\epsilon_i - \epsilon_a} 
  \label{Eq:2nd_cSE_1}
\end{eqnarray}
where $\hat{f}$ is the single-particle HF Hamiltonian, or simply Fock operator. Thus Eq.~(\ref{Eq:SE}) in the
main paper is derived.

For the HF reference state, i.e., when $\hat{v}^\text{MF} = \hat{v}^\text{HF}$, the $\psi$'s are eigenstates of the Fock operator, and hence Eq.~(\ref{Eq:SE_expression}) is zero. For any other reference state, e.g., a KS reference state, the  $\psi$'s are no longer eigenstates of the Fock operator, and Eq.~(\ref{Eq:SE_expression}) is in general not zero. This gives rise to a finite SE contribution to the second-order correlation energy.

\section{\label{sec:app/rSE}Derivation of the renormalized single excitation (rSE) contribution}

We start with the expression for the second-order single-excitation (SE) contribution 
discussed in Appendix~\ref{app:SE_deriv}
 \begin{equation}
    E_\text{c}^\text{SE} = \sum_{i,a} \frac{\langle \Phi_0 | \hat{H'}| 
             \Phi_{i}^{a}\rangle \langle \Phi_{i}^{a}| \hat{H'} |\Psi_0 \rangle }{E^{(0)}_0 - E^{(0)}_{ia}} \, .
    \label{Eq:E_SE_original}
 \end{equation}
The form of this equation actually already implies that the singly excited states 
$|\Phi_{i}^{a} \rangle$ are Slater determinants composed of \emph{canonical} orbitals, 
namely $|\Psi_{i}^{a}\rangle=\text{Det}\left\{ \psi_q \right\}$ where $\hat{h}^0 |\psi_q\rangle = 
\epsilon_q |\psi_q\rangle$, and $\hat{H}_0|\Phi_{i}^{a}\rangle = E^{(0)}_{ia}|\Phi_{i}^{a}\rangle$ with
$E^{(0)}_{ia}=E^{(0)}_0+\epsilon_a-\epsilon_i$. Appendix~\ref{app:SE_deriv} shows that Eq.~(\ref{Eq:E_SE_original}) 
can be reduced to the simple expression in Eq~(\ref{Eq:2nd_cSE_1})  that is given in terms of (\emph{canonical}) 
single-particle orbitals.

To set the stage for later discussions, we can also more generally express the SE energy in Eq.~(\ref{Eq:E_SE_original}) in terms of \emph{non-canonical} orbitals $\{\chi_q\}$, where $\hat{h}^0|\chi_p\rangle=\sum_{q}h^0_{pq}|\chi_q\rangle$, and
$h^0_{pq}=\langle \chi_p | \hat{h}^0| \chi_q\rangle$.  In this case, $E_\text{c}^\text{SE}$ is given by
 \begin{align}
  E_\text{c}^\text{SE} & = \sum_{ij,ab} \langle \Phi_0 | \hat{H'}| 
             \Phi_{i}^{a}\rangle \langle \Phi_{i}^{a} |(E^{(0)}_0 - \hat{H}_0)^{-1}|\Phi_{j}^{b} \rangle 
             \langle \Phi_{j}^{b}| \hat{H'} |\Phi_0 \rangle  \nonumber \\
         & = \sum_{ij,ab} \langle \chi_i | \hat{f} |\chi_a \rangle
         \left[(E^{(0)}_0 I - H_0)^{-1}\right]_{ia,jb}\langle \chi_b | \hat{f} |\chi_j\rangle \, ,
  \label{Eq:E_SE_noncanonical}
 \end{align}
where $I$ is the identity matrix: $I_{ia,jb} = \delta_{ij} \delta_{ab}$, and
  \begin{align}
    \left[E^{(0)}_0 I-H_0\right]_{ia,jb} & = \langle \Phi_{i}^{a} | E^{(0)}_0-\hat{H}_0 | \Phi_{j}^{b} \rangle \nonumber \\
       & = h^0_{ij} \delta_{ab} - h^0_{ab} \delta_{ij} \, .
    \label{Eq:SE_denominator}
  \end{align}

Now the question arises how to sum up all the higher-order SE diagrams shown in  Fig.~\ref{Fig:rSE_diagrams}?
For \emph{canonical} orbitals, the corresponding algebraic expression can be
easily obtained by applying the rules of evaluating Goldstone diagrams.\cite{Szabo/Ostlund:1989}
 \begin{widetext}
  \begin{align}
   E_\text{c}^\text{rSE} = & \sum_{ia}\frac{f_{ai}f_{ia}}{\epsilon_i - \epsilon_a} -
      \sum_{ij,a}\frac{f_{ai}\Delta v_{ij} f_{ja}}{(\epsilon_i-\epsilon_a)(\epsilon_j-\epsilon_a)} +
      \sum_{i,ab}\frac{f_{ai}f_{ib}\Delta v_{ba}}{(\epsilon_i-\epsilon_a)(\epsilon_i-\epsilon_b)} \nonumber \\ 
   + & \sum_{ijk,a}\frac{f_{ai}\Delta v_{ik} \Delta v_{kj} f_{ja}}
        {(\epsilon_i-\epsilon_a)(\epsilon_k-\epsilon_a)(\epsilon_j-\epsilon_a)} + 
      \sum_{i,abc}\frac{f_{ai}f_{ib}\Delta v_{bc} \Delta v_{ca}}
        {(\epsilon_i-\epsilon_a)(\epsilon_i-\epsilon_b)(\epsilon_i-\epsilon_c)}   \nonumber \\
   -  & \sum_{ij,ab}\frac{f_{ai}\Delta v_{ij} f_{jb} \Delta v_{ba}}
        {(\epsilon_i-\epsilon_a)(\epsilon_j-\epsilon_a)(\epsilon_j-\epsilon_b)} - 
     \sum_{ij,ab}\frac{f_{ai}\Delta v_{ij} f_{jb}\Delta v_{ba} }
        {(\epsilon_i-\epsilon_a)(\epsilon_i-\epsilon_b)(\epsilon_j-\epsilon_b)}  \\
 + &  \cdots  \nonumber
   \label{Eq:rSE_canonical}
  \end{align}
 \end{widetext}
where $\Delta v_{pq}=\langle \psi_p|\hat{f}-\hat{h}_0|\psi_q\rangle$, and 
$\Delta v_{ia} = f_{ia} = \langle \psi_i|\hat{f} |\psi_a\rangle$.
To see how the infinite-order summation in Eq~(\ref{Eq:rSE_canonical}) is carried out, 
we rearrange the expression as follows:
 \begin{widetext}
 \begin{align}
      E_\text{c}^\text{rSE} = & \sum_{ij,ab}\frac{f_{ai}\delta_{ij}\delta_{ab}f_{jb}}{\epsilon_i - \epsilon_a} +
      \sum_{ij,ab}\frac{f_{ai}\left(\Delta v_{ab}\delta_{ij} - \Delta v_{ij}\delta_{ab} \right)f_{jb}}
        {(\epsilon_i-\epsilon_a)(\epsilon_j-\epsilon_b)}  +\nonumber \\
      & \sum_{ijk,abc}\frac{f_{ai}\left(\Delta v_{ik} \Delta v_{kj} \delta_{ac} \delta_{bc} + 
          \Delta v_{bc} \Delta v_{ca} \delta_{ik} \delta_{kj} -
          \Delta v_{ik} \Delta v_{bc} \delta_{jk} \delta_{ac} -
          \Delta v_{kj} \Delta v_{ca} \delta_{ik} \delta_{bc} \right)f_{jb}}
        {(\epsilon_i-\epsilon_a)(\epsilon_k-\epsilon_c)(\epsilon_j-\epsilon_b)}   + \cdots \nonumber \\
      = &  \sum_{ij,ab}\frac{f_{ai}}{\epsilon_i - \epsilon_a} 
          \left[\delta_{ij}\delta_{ab} + \Omega_{ia,jb} + (\Omega^2)_{ia,jb} + \cdots \right] f_{jb} \nonumber \\
      = &
         \sum_{ij,ab}\frac{f_{ai}}{\epsilon_i - \epsilon_a} \left[ (I - \Omega)^{-1} \right]_{ia,jb} f_{jb}
 \end{align}
 \end{widetext}
where we have introduced the $\Omega$ matrix, defined as
 \begin{equation}
    \Omega_{ia,jb} = \frac{\Delta v_{ba}\delta_{ij}-\Delta v_{ij} \delta_{ab}}{\epsilon_j - \epsilon_b} \, .
 \end{equation}
Further denoting $A_{ia,jb}=(\epsilon_i - \epsilon_a)\delta_{ij}\delta_{ab}$, one observes
 \begin{align}
    \frac{1}{\epsilon_i-\epsilon_a}\left[(I-\Omega)^{-1}\right]_{ia,jb} 
  & = \left[A^{-1}((I-\Omega)^{-1}\right]_{ia,jb} \nonumber \\
  & =\left[(A-\Omega A)^{-1}\right]_{ia,jb} \, ,
 \end{align}
and
  \begin{align}
     (A-\Omega A)_{ia,jb} & = (\epsilon_i - \epsilon_a)\delta_{ij}\delta_{ab} + \Delta v_{ij} \delta_{ab} -
              \Delta v_{ba}\delta_{ij} \nonumber \\
      & = f_{ij}\delta_{ab} - f_{ab}\delta_{ij}  \, ,
     \label{Eq:rSE_denominator}
  \end{align}
where $f_{ij} = \epsilon_i \delta_{ij} + \Delta v_{ij}$,
$f_{ab}  = \epsilon_a \delta_{ab} + \Delta v_{ab}$ have been used. It follows that
  \begin{equation}
      E_\text{c}^\text{rSE} = \sum_{ij,ab} f_{ai}\left[(A-\Omega A)^{-1}\right]_{ia,jb}f_{jb} \, .
     \label{Eq:rSE_reduced}
  \end{equation}
We observe that the rSE energy expressed in terms of \textit{canonical} orbitals via 
Eqs.~(\ref{Eq:rSE_denominator}) and (\ref{Eq:rSE_reduced})
has the same mathematical structure as the second-order SE energy expressed in terms of 
\emph{non-canonical} orbitals given by
Eq.~(\ref{Eq:E_SE_noncanonical}) and (\ref{Eq:SE_denominator}). The difference is that
now the corresponding matrix elements in the denominator are evaluated using the Fock
operator $\hat{f}$, instead of the KS Hamiltonian operator $\hat{h}^0$.

To simplify the evaluation of Eq.~(\ref{Eq:rSE_reduced}), one can rotate
the occupied orbitals and unoccupied orbitals separately, such that the Fock matrix becomes
diagonal in the occupied and unoccupied subspaces. This procedure is called \emph{semi-canonicalization}. 
To be more precise, suppose there
are transformation matrices ${\cal O}$ and ${\cal U}$ which diagonalize the $f_{ij}$ and
$f_{ab}$ blocks separately
  \begin{align}
     \sum_{k} f_{ik} {\cal O}_{kj}= {\cal O}_{ij}\tilde{\epsilon}_j \nonumber \\
     \sum_{c} f_{ac} {\cal U}_{cb}= {\cal U}_{ab}\tilde{\epsilon}_b  \, .
    \label{Eq:f_block_rotation}
  \end{align}
We then have
  \begin{equation}
       \sum_{kl,cd} {\cal O}^\ast_{ik}{\cal U}^\ast_{ac} (A-\Omega A)_{kc,ld} {\cal O}_{lj}{\cal U}_{db} = 
       \delta_{ij}\delta_{ab}(\tilde{\epsilon}_j - \tilde{\epsilon}_b)  \,
  \end{equation}
or equivalently,
  \begin{equation}
       \left[(A-\Omega A)^{-1}\right]_{ia,jb} = 
      \sum_{k,c} {\cal O}_{ik}{\cal U}_{ac}(\tilde{\epsilon}_k - \tilde{\epsilon}_c)^{-1} 
      {\cal O}_{kj}^\ast {\cal U}_{cb}^\ast  \, .
   \label{Eq:rSE_denom_transform}
  \end{equation}
Inserting Eq.~(\ref{Eq:rSE_denom_transform}) into Eq.~(\ref{Eq:rSE_reduced}), one arrives at
  \begin{equation}
      E_\text{c}^\text{rSE} = \sum_{ia} \frac{\tilde{f}_{ai}\tilde{f}_{ia}}{\tilde{\epsilon}_i - \tilde{\epsilon}_a} \, .
     \label{Eq:rSE_final}
  \end{equation}
where 
 \begin{equation}
   \tilde{f}_{ia} = \sum_{jb} {\cal O^\ast}_{ij}{\cal U^\ast}_{ab} f_{jb} \, .
   \label{Eq:amplitude_transform}
 \end{equation}
Thus the final expression for rSE has the same form as that for SE, only the eigenvalues 
$\epsilon_i,\epsilon_a$ and the ``transition amplitude" $f_{ia}$ have to be reinterpreted. The actual 
implementation following Eqs.~(\ref{Eq:f_block_rotation}), (\ref{Eq:rSE_final}), and 
(\ref{Eq:amplitude_transform}) is straightforward.

\section{\label{sec:app/basis}Basis convergence}
\begin{figure}
\includegraphics[width=0.45\textwidth]{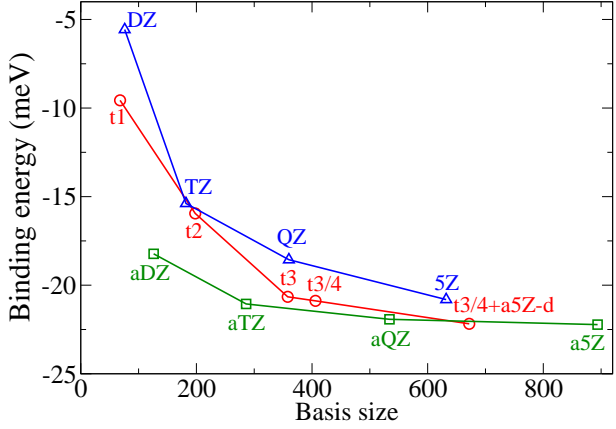}
\caption{ 
 (Color online) 
The rPT2@PBE binding energy of methane dimer in its equilibrium geometry as a function of the basis set size.
``XZ" and ``aXZ" (X=D,T,Q,5) denote respectively the Dunning ``cc-pVXZ" and ``aug-cc-pVXZ" basis, whereas
``tN" denotes the FHI-aims ``\textit{tier} N" basis, and ``t3/4" here  means \textit{tier} 4 basis for C 
and \textit{tier} 3 basis for H (note that a \textit{tier} 4 basis for H is not available). 
``t3/4+a5Z-d" corresponds to the NAO ``\textit{tier} 3/4" plus diffuse functions from aug-cc-pV5Z. 
The BSSE is corrected.}

\label{Fig:methane_dimer}
\end{figure}

Figure~\ref{Fig:Ar2_rSE-d} shows the convergence behavior of the rPT2 binding energy of the methane dimer (in
its equilibrium geometry) with respect to the FHI-aims NAO ``\textit{tier} N" basis as well as Dunning's
``cc-pVXZ" and  ``aug-cc-pVXZ" basis. The methane dimer is dominated by the dispersion interaction, and
the so-called ``diffuse functions"  are needed to accurately describe this interaction. The difference between
the ``cc-pVXZ" and ``aug-cc-pVXZ" results highlight the  importance of including ``diffuse functions".
For methane dimer  the ``\textit{tier} N" series exhibits a faster convergence than ``cc-pVXZ" whereas a slower 
convergence than ``aug-cc-pVXZ" for BSSE-corrected binding energies. When
adding diffuse functions from aug-cc-pV5Z to ``\textit{tier} 3/4", (called
``t3/4+a5Z-d" in Fig.~\ref{Fig:methane_dimer}) results of similar quality as the full aug-cc-pV5Z basis are 
obtained.

\end{appendix}
\bibliography{./CommonBib}
\end{document}